\PassOptionsToPackage{table}{xcolor}

\documentclass[sigconf]{acmart}
\usepackage{array}
\usepackage{longtable} 
\usepackage{verbatimbox} 
\usepackage{fancyvrb} 
\usepackage{multicol}

\usepackage{float} 
\usepackage{graphicx} 
\usepackage{float} 
\usepackage{tabularx} 
\usepackage{booktabs} 
\usepackage{comment}
\usepackage{marvosym}
\usepackage{ragged2e}
\usepackage[utf8]{inputenc} 

\usepackage{pifont}
\newcommand{\xmark}{\ding{55}}%

\usepackage{balance} 

\usepackage[table]{xcolor}
\usepackage[colorinlistoftodos]{todonotes} 
\usepackage{enumitem}
\AtBeginDocument{%
  }

\copyrightyear{2025}
\acmYear{2025}
\setcopyright{cc}
\setcctype{by}
\acmConference[KDD '25]{Proceedings of the 31st ACM SIGKDD Conference on Knowledge Discovery and Data Mining V.2}{August 3--7, 2025}{Toronto, ON, Canada}
\acmBooktitle{Proceedings of the 31st ACM SIGKDD Conference on Knowledge Discovery and Data Mining V.2 (KDD '25), August 3--7, 2025, Toronto, ON, Canada}
\acmDOI{10.1145/3711896.3737417}
\acmISBN{979-8-4007-1454-2/2025/08}

\begin{document}
\newcommand{\datasetname}{\texttt{VideoConviction}}

\author{Michael Galarnyk}
\authornote{Authors contributed equally to this research.}
\affiliation{%
  \institution{Georgia Institute of Technology}
  \city{Atlanta}
  \state{GA}
  \country{USA}
}

\title{VideoConviction: A Multimodal Benchmark for Human Conviction and Stock Market Recommendations}

\author{Veer Kejriwal$^{*}$}
\affiliation{%
 \institution{Georgia Institute of Technology}
 \city{Atlanta}
  \state{GA}
  \country{USA}
  }

\author{Agam Shah$^{*}$ \Letter}

\affiliation{%
 \institution{Georgia Institute of Technology}
 \city{Atlanta}
 \state{GA}
 \country{USA}}

\author{Yash Bhardwaj}
\authornote{All work done as a Volunteer Research Assistant at Georgia Institute of Technology. \\\Letter: \href{mailto:ashah482@gatech.edu}{ashah482@gatech.edu}}
\affiliation{%
 \institution{Georgia Institute of Technology}
 \city{Atlanta}
  \state{GA}
  \country{USA}
}

\author{Nicholas Watney Meyer}
\affiliation{%
 \institution{Georgia Institute of Technology}
 \city{Atlanta}
  \state{GA}
  \country{USA}
}

\author{Anand Krishnan$^\dagger$}
\affiliation{%
 \institution{Stanford University}
 \city{Palo Alto}
  \state{CA}
  \country{USA}
} 

\author{Sudheer Chava}
\affiliation{%
 \institution{Georgia Institute of Technology}
 \city{Atlanta}
  \state{GA}
  \country{USA}
}

\renewcommand{\shortauthors}{Michael Galarnyk, Veer Kejriwal, Agam Shah, et al.}

\begin{abstract}
Social media has amplified the reach of financial influencers known as "finfluencers," who share stock recommendations on platforms like YouTube. Understanding their influence requires analyzing multimodal signals like tone, delivery style, and facial expressions, which extend beyond text-based financial analysis. We introduce \textbf{VideoConviction}, a multimodal dataset with 6,000+ expert annotations, produced through 457 hours of human effort, to benchmark multimodal large language models (MLLMs) and text-based large language models (LLMs) in financial discourse. Our results show that while multimodal inputs improve stock ticker extraction (e.g., extracting Apple's ticker AAPL), both MLLMs and LLMs struggle to distinguish investment actions and conviction—the strength of belief conveyed through confident delivery and detailed reasoning—often misclassifying general commentary as definitive recommendations. While high-conviction recommendations perform better than low-conviction ones, they still underperform the popular S\&P 500 index fund. An inverse strategy—betting against finfluencer recommendations—outperforms the S\&P 500 by 6.8\% in annual returns but carries greater risk (Sharpe ratio of 0.41 vs. 0.65). Our benchmark enables a diverse evaluation of multimodal tasks, comparing model performance on both full video and segmented video inputs. This enables deeper advancements in multimodal financial research. Our \href{https://github.com/gtfintechlab/VideoConviction}{\textcolor{blue}{code}}, \href{https://huggingface.co/datasets/gtfintechlab/VideoConviction}{\textcolor{blue}{dataset}}, and \href{https://huggingface.co/spaces/gtfintechlab/VideoConvictionLeaderboard}{\textcolor{blue}{evaluation leaderboard}} are available under the CC BY-NC 4.0 license.
\end{abstract}




\maketitle


\section{Introduction}
Social media has given rise to finfluencers, individuals who provide stock market and investment recommendations on platforms like YouTube and Instagram \cite{Guan_2023}. According to the Securities and Exchange Commission (SEC), a recommendation is any communication that acts as a “call to action” or “reasonably would influence an investor to trade a particular security or group of securities” \cite{SEC_RegBI}. With their widespread reach, finfluencers shape retail investors’ (individual, non-professional traders) financial decisions, influencing how they assess and respond to market opportunities \cite{Symbiosis_Gandhi_2024}. Instead of solely focusing on financial returns, finfluencers often monetize their influence through social media engagement and affiliations with trading platforms \cite{Hull_Qi_2024b}. 

The evolving incentives of finfluencers have raised questions about social media’s impact on financial markets. Technological advancements and social media have been argued to reduce stock market efficiency by increasing noise and encouraging short-term speculative trading driven by persuasive and sensationalized content \cite{asness2024less,WARKULAT2024103721}. An analysis of finfluencer tweets revealed that 56\% consistently gave advice leading to poor financial outcomes. Despite this, these same finfluencers attracted larger followings and exerted more influence on retail investors than those who provided more reliable financial guidance \cite{kakhbod2023finfluencers}.

\begin{table*}[ht]
\footnotesize
\centering
    \renewcommand{\arraystretch}{1.1}
\caption{Comparison of datasets across multiple attributes. \textit{Source} indicates the source of raw data. \textit{Modalities} indicate data types—Visual (V), Audio (A), and Text (T). \textit{Domain} specifies the dataset’s focus area. \textit{Duration} is total video length. \textit{\# Videos} is the total number of videos. \textit{Transcripts/ASR} indicates transcript availability. \textit{Conviction} specifies if conviction labels are present. \textit{Annotator Type} describes annotators’ background. Notably, \datasetname{} is the first multimodal dataset of its kind in a domain-specific area (Finance), featuring expert annotation, transcripts/ASR, and a multimodal conviction score.}
\label{tab:comparison_datasets}
\resizebox{1.0\textwidth}{!}{%
\begin{tabular}{lcccccccc}
\toprule
\textbf{Dataset} & \textbf{Source} & \textbf{Modalities} & \textbf{Domain}  
& \textbf{Duration (hours)} & \textbf{\# Videos} & \textbf{Transcripts/ASR} & \textbf{Conviction} & \textbf{Annotator Type} \\ 
\midrule

\textbf{ExFunTube} \citep{ko2023can} &
YouTube & [V, A, T] & HCI & 84 & 10K & \checkmark & \xmark & Amazon Mechanical Turk \\ 

\textbf{DeVan} \citep{liu2024devandensevideoannotation} &
YouTube & [V, A, T] & Misc. & 96 & 6.7K & \checkmark & \xmark & College and Graduate-level Students \\ 

\textbf{YouTube-8M} \citep{abuelhaija2016youtube8mlargescalevideoclassification} &
YouTube & [V, A, T] & Misc. & 500K & 8M & \xmark & \xmark & Automated \\ 

\textbf{ActivityNet} \citep{caba2015activitynet} &
YouTube & [V, A, T] & Household & 849 & 27.8K & \xmark & \xmark & Amazon Mechanical Turk \\ 

\textbf{HowTo100M} \citep{miech19howto100m} &
YouTube & [V, A, T] & Misc. & 134K & 1.221M & \checkmark & \xmark & Automated \\ 

\textbf{TweetFinSent} \citep{pei-etal-2022-tweetfinsent} &
Twitter & [T] & Finance & NA & NA & \xmark & \xmark & Expert Humans \\ 

\textbf{FOMC} \citep{shah-etal-2023-trillion} &
Federal Reserve, USA  & [T] & Finance & NA & NA & \checkmark & \xmark & Expert Humans \\ 

\textbf{HourVideo} \citep{chandrasegaran2024hourvideo1hourvideolanguageunderstanding} &
Ego4D \citep{grauman2022ego4dworld3000hours} & [V, A, T] & HCI & 381 & 500 & \xmark & \xmark & Expert Humans \\ 

\textbf{CMU-MOSEI} \citep{bagher-zadeh-etal-2018-multimodal} &
YouTube & [V, A, T] & HCI & 66 & 3.2K & \checkmark & \xmark & Amazon Mechanical Turk \\

\midrule
\textbf{\datasetname{} } &
YouTube & [V, A, T] & Finance & 43 & 288 & \checkmark & \checkmark & Expert Humans \\ 

\bottomrule
\end{tabular}
}
\end{table*}

A key question, then, is why investors continue to follow finfluencers whose recommendations lead to poor financial outcomes. A possible explanation is that underperforming influencers present their recommendations with greater conviction—a deeply held belief reinforced by emotional intensity and detailed reasoning \cite{Abelson}. Conviction, combined with overly positive sentiment, makes retail investors more likely to follow poor investment advice \cite{kakhbod2023finfluencers}. However, most studies on finfluencers have focused on text \cite{sprengertweets2014, GrossKlussmann2019}, overlooking how multimodal media impacts audience perception, influence emotions, and contribute to their persuasive impact \cite{10.1093/jcmc/zmab010, 10.1145/3665026.3665039}.

To address these gaps, we introduce \datasetname\ , a novel, expert-annotated multimodal dataset focused on financial discourse, derived from YouTube videos of finfluencers discussing the US stock market. Unlike existing datasets (see Table \ref{tab:comparison_datasets}), \datasetname\  is a domain-specific financial dataset with expert annotations and multimodal content. The dataset serves as a benchmark for multimodal models to capture nuances in human communication, such as conviction, by leveraging audiovisual signals like tone, facial expressions, and gestures. It includes both \textbf{full-length video URLs with corresponding transcripts} and \textbf{segmented videos with corresponding transcripts}, allowing models to analyze recommendations at different granularities. Table \ref{tab:dataset_metrics} provides an overview of dataset metrics. 

\begin{table}[H]
    \centering
    \renewcommand{\arraystretch}{0.85} 
    \setlength{\tabcolsep}{6pt} 
    \caption{Overview of the \datasetname\ dataset.}
    \label{tab:dataset_metrics}
    \begin{tabular}{l c}
        \toprule
        \textbf{Metric} & \textbf{Value} \\
        \midrule
        Total Full Length Videos & 288 \\
        Total Video Segments & 687 \\
        Total Duration (hours) & 43 \\
        Expert Annotation Effort (hours) & 457 \\
        Avg. Video Length (minutes) & 9 \\ 
        Total Annotated Datapoints & 6,063 \\
        Total Datapoints & 23,278 \\
        Years Covered & 2018--2024 \\
        \bottomrule
    \end{tabular}
\end{table}

Beyond introducing this dataset, our study examines the broader implications of financial influencers on social media, particularly their role in shaping retail investor behavior and market dynamics. By linking multimodal signals to measurable outcomes (e.g., stock market reactions), our dataset enables models to better understand the interplay between verbal and non-verbal communication in financial discourse. Our contributions are summarized as follows:

\begin{itemize}
    \item \datasetname\ comprises 6,000+ expert-annotations, capturing multimodal conviction through segmented videos.

    \item We benchmark MLLMs and LLMs on full-length and segmented content to assess the impact of multimodalities, identifying when non-textual signals improve model performance.

    \item We demonstrate a significant gap between models and human-level understanding of informal, multimodal dialogue found in social media content.
    
    \item Our portfolio analysis reveals that while high-conviction influencer recommendations outperform low-conviction ones, they still underperform market benchmarks. An inverse strategy—betting against finfluencer recommendations—yields higher returns but comes with increased risk.

\end{itemize}

\section{Related Work}

\subsection{Multimodal Datasets}
A detailed comparison of existing multimodal benchmarks with \datasetname\ is presented in Table \ref{tab:comparison_datasets}. Most prior datasets focus on human-object interactions in videos \cite{miech19howto100m, krishna2017densecaptioningeventsvideos} or question-answering tasks \cite{liu2024devandensevideoannotation, chandrasegaran2024hourvideo1hourvideolanguageunderstanding}. Some of them \cite{abuelhaija2016youtube8mlargescalevideoclassification, bagher-zadeh-etal-2018-multimodal, ko2023can}, annotate sentiment and emotion but assess only conveyed emotions, not the strength of conviction. Despite the growing influence of social media, little research evaluates multimodal content from influencers. Current MLLMs \cite{chandrasegaran2024hourvideo1hourvideolanguageunderstanding} focus on physical interactions, overlooking the complex ways social media personalities shape viewer perceptions. 

Compared to existing datasets, \datasetname\ has a smaller total video duration but offers highly granular, expert annotations, including a novel conviction score, applied at precise time windows rather than at the whole video level. It extends prior research on retail trader sentiment \cite{pei-etal-2022-tweetfinsent, liang2024enhancingfinancialmarketpredictions}, which lacks multimodal components, addressing the increasing prominence of audiovisual content in financial communication. Unlike \citet{chandrasegaran2024hourvideo1hourvideolanguageunderstanding}, this dataset explicitly tracks human interaction within social media contexts while benchmarking multiple tasks, as shown in Table \ref{tab:comparison_datasets}. 

Existing benchmarks evaluate video understanding, prior knowledge incorporation, and decision-making  \cite{ning2023videobenchcomprehensivebenchmarktoolkit}, but MLLMs continue to fall short of human-like comprehension, as confirmed by our study. Also, widely used benchmarks \cite{liu2024mmbenchmultimodalmodelallaround} fail to measure conviction interpretation, a key aspect of financial communication. By addressing this gap, \datasetname\ aims to advance research on MLLMs' ability to analyze convincing financial content.

\begin{figure*}[ht]
    \centering
    \includegraphics[width=\textwidth]{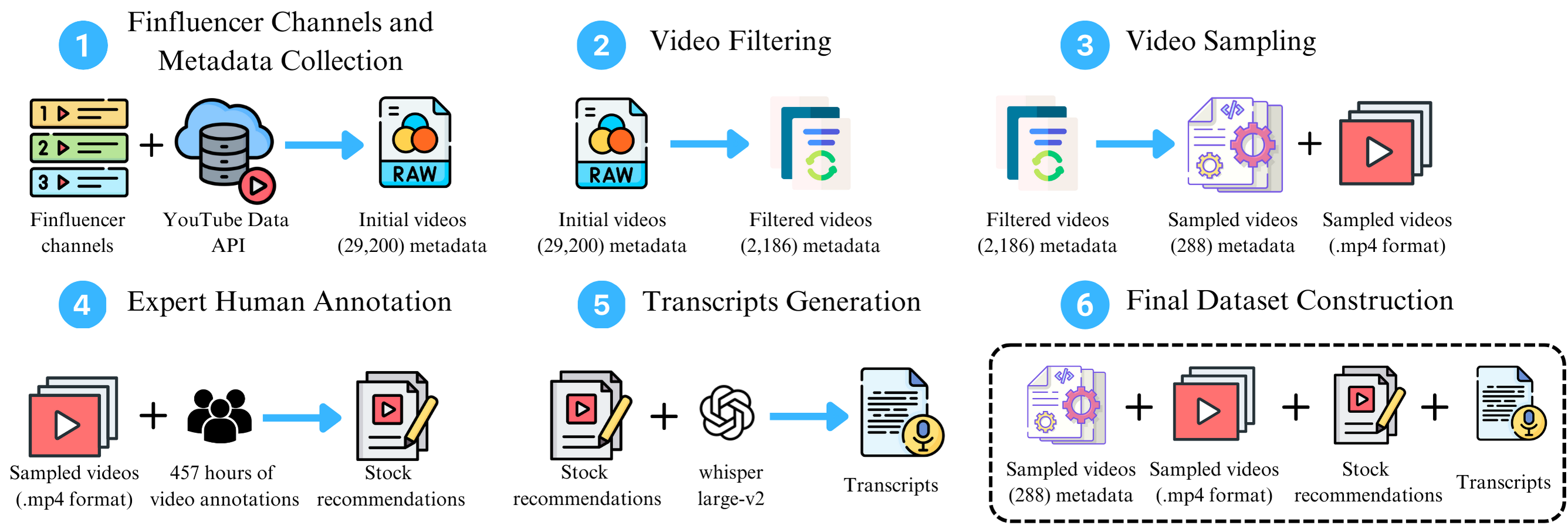}
    \caption{The VideoConviction dataset generation pipeline consists of six stages, combining automated tools and expert human annotations to ensure high-quality stock recommendation data. (1) We curate a list of 22 finfluencer channels and collect metadata for 29,200 videos using the YouTube Data API. (2) Metadata-based filtering reduces this set to 2,186 videos. (3) A final sample of 288 videos is selected and downloaded in .mp4 format. (4) Expert annotators spent 457 hours to label stock recommendations and assign conviction scores. (5) Automatic speech recognition (ASR) generates transcripts. (6) The final VideoConviction dataset integrates metadata, annotations, and transcripts.}
    \label{fig:dataset_pipeline}
\end{figure*}

\subsection{AI in Finance}
\paragraph{Multimodality in Finance}
Multimodal sentiment analysis combines textual, visual, and acoustic data to improve sentiment prediction \cite{ZHU2023306,DasSingh2023}. In finance, earnings calls integrate verbal and vocal cues with financial features, using graph convolutions to model stock interdependencies \cite{sawhney-etal-2020-voltage}. Beyond audiovisual signals, approaches like GAME \cite{Ang2022Guided} fuse numerical data with global and local news to improve stock return and volatility forecasting. Integrating cross-market financial data via multimodal deep learning enhances market prediction \cite{Lee2020}. Challenges remain in aligning asynchronous modalities, handling financial jargon, and adapting to linguistic and cultural variations. Insights from multimodal sentiment analysis aid market sentiment and risk assessment \cite{ZHU2023306}.

\paragraph{Social Media and Retail Investors}
Social media platforms like YouTube and TikTok amplify nontraditional finance, influencing retail investors through engaging and persuasive content \cite{kakhbod2023finfluencers}. “Finfluencers” attract large audiences, often promoting unreliable advice, yet their delivery style—body language, vocal tone, and presentation—significantly shapes market sentiment \cite{asness2024less}. Unlike text-based sentiment analysis, which overlooks nonverbal cues, multimodal research highlights how high-energy videos can drive impulsive investor behavior, sometimes fueling market inefficiencies \cite{10.1093/rapstu/rax020, de2022young, chacon2023will}. To capture these effects, computational linguistics must extend beyond text, incorporating visual and vocal signals to study credibility-building techniques like eye contact, confident posture, and emotive speech. Expanding sentiment analysis in this direction improves understanding of social media’s role in shaping market trends \cite{kakhbod2023finfluencers}.

\paragraph{LLMs \& Benchmarks on Financial Data}
Financial LMs have evolved since FinBERT \cite{araci2019finbert}, one of the first domain-specific models, designed for sentiment analysis. Subsequent models, including FLANG \cite{shah2022flue}, BloombergGPT \cite{wu2023bloomberggpt}, FinGPT \cite{yang2023fingpt}, and BBT-Fin \cite{lu2023bbt}, refine financial NLP via diverse training data and architectures. These LMs primarily rely on formal financial sources like earnings calls, 10-K filings, and media reports, leveraging datasets such as FiQA \citep{FiQA}, Sentiment Analysis \citep{malo2013gooddebtbaddebt}, FOMC \citep{shah-etal-2023-trillion}, and World Central Banks \citep{shah2025wordsuniteworldunified}. Early sentiment analysis efforts used Loughran and McDonald's \citep{loughran2011liability} financial dictionary, which classified sentiment as positive or negative. Recent work, such as SubjECTive-QA \citep{pardawala2024subjectiveqameasuringsubjectivityearnings}, expands sentiment modalities to six, capturing nuanced corporate financial language. 

Notably, existing datasets remain text-based, despite multimodal finance media’s rise. \datasetname\ addresses this gap and focuses on informal financial discourse, an underexplored area. 

\section{\datasetname\ Dataset}
\label{sec:dataset}

\datasetname\ is a curated dataset of 288 finfluencer videos, segmented into 673 expert-annotated recommendations. Each segment includes key details—ticker name, action, action source—and a multimodal conviction score. Along with annotations, we provide full video URLs, timestamps of segments, ASR transcripts, and rich metadata for downstream analysis. The dataset includes 6,063 expert-annotated gold labels and 23,278 total data points, spanning a diverse set of stock recommendations with 7M+ YouTube views and a 104M+ audience. \datasetname\ enables three key applications: (1) MLLM benchmarking, (2) LLM benchmarking, and (3) financial portfolio analysis via backtesting. We open-source the dataset construction pipeline, which is summarized in Figure \ref{fig:dataset_pipeline}.

\subsection{Finfluencer Channel Curation}
\label{sec:finfluencer_channel_curation}
We manually curated a list of 22 finfluencer YouTube channels, referencing an initial selection from this article\footnote{\url{https://finance.yahoo.com/news/20-best-value-investing-youtube-111203385.html}}. To ensure temporal diversity, we included channels active between 2018 and 2024, capturing heterogeneous market conditions before, after, and during the pandemic surge in the U.S. stock market and the concurrent rise of finfluencers giving recommendations on YouTube. These channels vary in reach and backgrounds (see Appendix \ref{sec:finfluencer_background}), with subscriber counts between \textbf{21K and 733K} and total channel views ranging from \textbf{1M to 120M}. We focused on channels providing U.S. stock recommendations, excluding those centered on general market discussions, financial education, or alternative investments. Using the YouTube API \footnote{\url{https://console.cloud.google.com/apis/library/youtube.googleapis.com}}, we retrieved 29,200 video URLs along with their titles and additional metadata, as detailed in Appendix \ref{app:metadata_details}.

\subsection{Video Filtering}
\paragraph{Keywords} We systematically filter and retain videos relevant to stock recommendations by analyzing each video title using a rule-based approach combined with regex pattern matching. Specifically, we define 4 categories of keywords, as outlined in Table \ref{tab:keyword_categories}. Many of these keywords are adapted from \citet{chacon2023will}.

\begin{table}[ht]
\centering
    \renewcommand{\arraystretch}{0.9} 
\caption{Keyword Categories for Rule-based Video Filtering}
\label{tab:keyword_categories}
\begin{tabular}{p{0.35\linewidth}|p{0.55\linewidth}}
\toprule
\textbf{Recommendation} & \textbf{Non-Recommendation}\\
\midrule
buy, buying, bought, sell, selling, sold, hold, holding, held, bullish, bearish
& 
analysis, market analysis, review, market review, crypto, cryptocurrency, bitcoin, btc, etherium, eth, altcoin, altcoins, estate\\
\toprule
\textbf{Security Descriptor} & \textbf{Security Types}\\
\midrule
best, top, worth, worst, tanking & 
stock, stocks, etf, etfs, company, companies \\
\bottomrule

\end{tabular}
\end{table}

\paragraph{Rule-based Filtering Algorithm} A title is approved if it contains at least one recommendation keyword. Otherwise, it is evaluated against a predefined regex pattern for stock-related content. This pattern requires a security descriptor and a security type, with acceptance granted if either condition holds:

\begin{enumerate}
    \item A security descriptor precedes a security type (e.g., \textit{``Best Tech Stocks''})  
    \item A security type precedes a security descriptor (e.g., \textit{``Stocks with Best Returns''})  
\end{enumerate}  

If neither condition is met, the video title is checked for non-recommendation keywords. The presence of any such keyword leads to rejection. If none of the approval or rejection conditions are satisfied, the title remains ambiguous and is rejected by default.

\paragraph{Additional Filtering Criterion} Videos exceeding 12 minutes in duration were removed. The rationale is outlined in Appendix  \ref{app: length_of_video_rationale}. 

This entire filtering process brought our videos down from 29,200 to 2,186 videos.

\subsection{Video Sampling}
The objective of this phase is to mitigate temporal bias by ensuring that videos are evenly distributed across years. After filtering, we obtained a total of 2,186 videos. From this set, we selected a stratified sample of 288 videos, which forms the final video selection for our dataset. To maintain uniformity across the dataset, we restrict video formats to \texttt{[.mp4, .m4a]} and prioritize resolutions in the following order: $720p > 480p > 1080p$.

\subsection{Expert Human Annotation}
\label{sub: expert_human_annotation}
Five expert annotators manually labeled 288 long-form ($\leq12$ minutes) YouTube videos featuring finfluencers' U.S. stock market recommendations. For each video, the annotation process consisted of three key steps:

\begin{enumerate}
    \item \textbf{Video Title Annotation:} The annotator labeled the video title, sourced from metadata collected in Section \ref{sec:finfluencer_channel_curation}, as it was a contributing factor in the subsequent conviction score evaluation.
    \item \textbf{Stock Recommendation Identification:} The annotator watched the full video and determined whether it contained a stock recommendation, labeling \textit{Yes} or \textit{No}.
    \item \textbf{Segment Annotation:} If \textit{No}, it was a false positive however the video was still kept in the final dataset to represent a sample of what a finfluencer video with no recommendation looks like. If \textit{Yes}, the annotator rewatched the video and annotated the segments. If $N$ recommendations were present, the video was divided into $N$ segments, each with start and end timestamps. Each segment was annotated along the 3 dimensions shown in Table \ref{tab:recommendation_details}.
\end{enumerate} 

\begin{figure*}[ht]
    \centering
    \includegraphics[width=\textwidth]{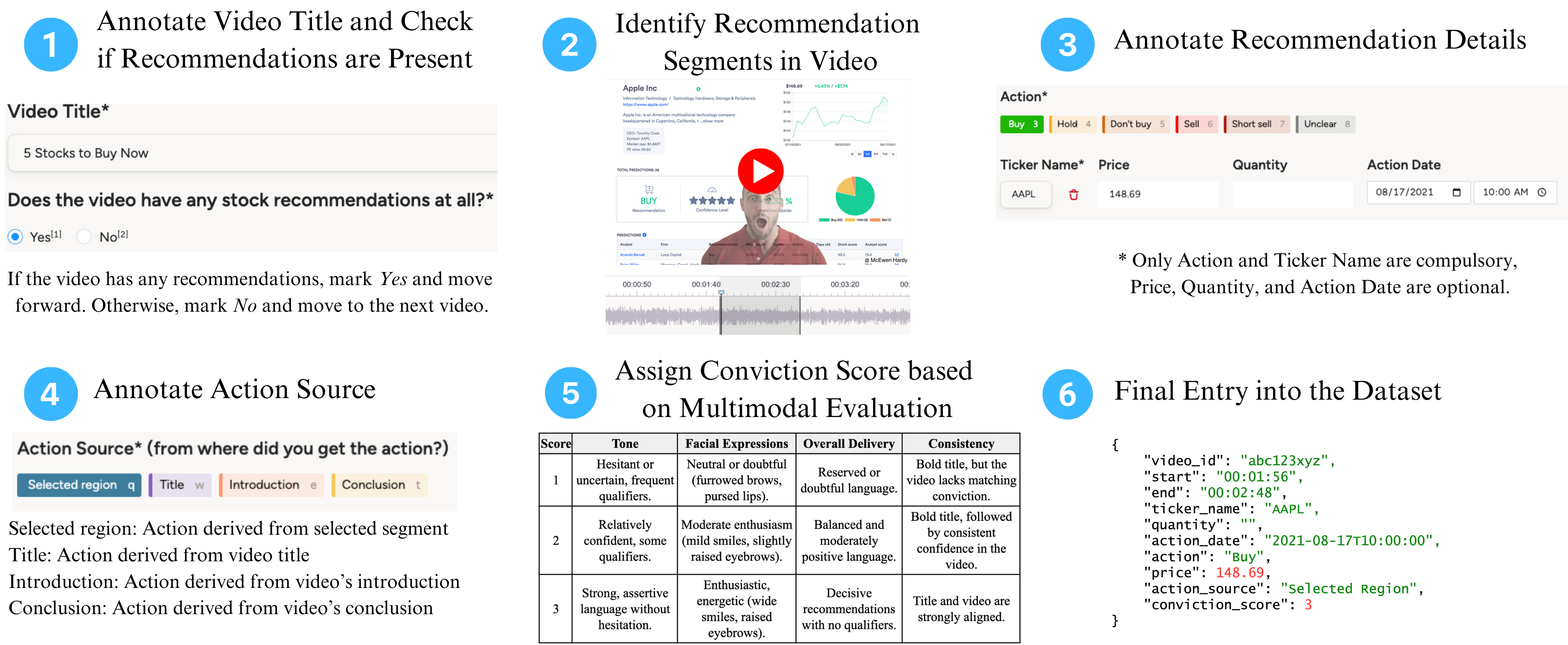}
    \caption{Step-by-step annotation workflow for extracting high-quality finfluencer recommendations from video content: (1) The annotator adds the video title and assesses if the video contains any stock recommendations at all. (2) If stock recommendations are present, relevant segments are selected. (3) Recommendation details are annotated. (4) The specific part of the video where the recommendation originates (action source) is identified. (5) A multimodal conviction score is assessed. (6) The final annotated entry, including all extracted details, is added to the dataset.}
    \label{fig:annotation_process}
    
\end{figure*}

\begin{table}[h]
\footnotesize
\centering
\renewcommand{\arraystretch}{1.2} 
\caption{Each segment is annotated across three dimensions: recommendation details, action source, and conviction score.}
\label{tab:recommendation_details}
\begin{tabular}{>{\raggedright\arraybackslash}p{0.25\linewidth}  >{\raggedright\arraybackslash}p{0.66\linewidth}}
\textbf{Recommendations Details} & 
\texttt{Ticker name} and \texttt{action} were compulsory fields, whereas \texttt{price}, \texttt{quantity}, and \texttt{action date} were annotated whenever available. See more details on how we defined a recommendation in Appendix~\ref{app:what-is-a-recommendation}. \\
\midrule
\textbf{Action Source} & 
Identifies the part of the video where the recommendation originates. If it appears in multiple regions, select \texttt{action source} in the following order: Selected Region > Title > Introduction > Conclusion. \\
\midrule
\textbf{Conviction Score} & 
Evaluate the fininfluencer’s \texttt{multimodal conviction score} (1–3, with 3 being the highest) in the selected region based on tone, facial expressions, overall delivery, and title-region consistency (see Table in Figure~\ref{fig:annotation_process}). We labeled \textbf{conviction}, following \citet{Abelson}. \\

\end{tabular}
\end{table}

A visual overview of the annotation process is presented in Figure \ref{fig:annotation_process}. We used Label Studio \cite{Label-Studio} for annotations. The labeling interface is shown in Appendix \ref{app:labeling_interface}. From 288 videos, we identified \textbf{687 unique stock recommendation segments}.  

\subsection{Transcript Generation (ASR)}
To generate video transcripts, we utilize Whisper-large model \cite{pmlr-v202-radford23a}, which outperforms YouTube Automatic Captions \cite{Rai2024NPTEL}. Transcripts are generated for both the full-length videos and individual recommendation segments, where each segment corresponds to the portion of the video between the specified start and end timestamps.

\subsection{Final Dataset Construction}
Each entry in the final \datasetname\ dataset represents a unique video segment tied to a single stock recommendation. While the actual column names of the dataset are different, the broad categories of data in the final dataset is shown in Table \ref{tab:dataset_structure}.

\begin{table}[H]
    \centering
    \caption{Brief description of data in \datasetname.}
    \label{tab:dataset_structure}
    \renewcommand{\arraystretch}{1.2}
    \begin{tabularx}{\linewidth}{>{\centering\bfseries}p{0.3\linewidth} >{\arraybackslash}X} 
        \toprule
        Video URL & YouTube link to the finfluencer video. \\ 
        Timestamps & Start and end times of each segment. \\ 
        Recommendation & Action, ticker, price, quantity, date. \\ 
        Action Source & Origin of the action. \\ 
        Conviction Score & Finfluencer's conviction when presenting recommendation. \\ 
        ASR Content & Transcript of full video and segment. \\ 
        Metadata & Video \& channel metadata (Appendix \ref{app:metadata_details}). \\ 
        \bottomrule
    \end{tabularx}
\end{table}

\paragraph{Data Integration} This final dataset was constructed by integrating three distinct data sources:

\begin{itemize}
    \item \textbf{Raw Annotations:} In the raw exported CSV, each row corresponded to a single video. Each column in that row represented a different feature (e.g., ticker name, action, conviction score). Because each video had been split into multiple segments during annotation, each cell (i.e., each row–column pair) contained a list of JSON objects—one JSON object per segment. Each JSON object captured the start timestamp, end timestamp, and labeled annotation for that particular feature in that specific segment.
    To transform these annotations into a final dataset, the lists of JSON objects were iterated over so that each segment (originally embedded in a list within a cell) became its own entry (row) in the resulting dataset, and each feature (column) remained distinct. This way, each row in the final dataset corresponds to one video segment, and each column provides the feature-specific annotation for that segment.

    \item \textbf{Transcripts/ASR:} For each segment, the corresponding transcript and the full video transcript are appended. This dual transcript approach preserves both the local context (segment-level) and the broader context (full-video), which is crucial for benchmarking and comparing the performance of MLLMs on whole videos vs. segmented videos.
    
    \item \textbf{Channel and Video Metadata:} Since each segment originates from a finfluencer video, the corresponding video and channel level metadata were integrated into every entry.

\end{itemize}

\paragraph{Data Validation \& Quality Check} We employed a hybrid quality assurance process, combining expert human review with automated validation. Five experts conducted peer reviews to ensure annotation consistency and accuracy. Additionally, a validation tool flagged missing fields, incorrect segment annotations, segment timestamp overlaps, and other inconsistencies. Errors were sent back for manual correction, followed by a revalidation cycle to ensure consistency. This iterative process upheld the high annotation standards, aligning with prior efforts \citep{chandrasegaran2024hourvideo1hourvideolanguageunderstanding, jin2024rjuameddqamultimodalbenchmarkmedical}.

\begin{table*}[ht]
    \centering
    \renewcommand{\arraystretch}{1.15}
    \setlength{\tabcolsep}{10pt}
    \caption{Benchmarking results (F1 scores) comparing model performance to human annotations across three tasks: (T) extracting the recommended stock ticker name, (TA) extracting the stock ticker name and its corresponding action, and (TAC) extracting the stock ticker name, action, and conviction score. Models are evaluated on full-length and segmented inputs. LLMs use transcripts only; MLLMs use both transcripts and video. Best scores in each column are highlighted in green.}
    \label{tab:benchmarking_results}
    
    \renewcommand{\arraystretch}{0.95} 
    \begin{tabular}{lcc|p{0.9cm}<{\centering} p{0.9cm}<{\centering} p{0.9cm}<{\centering}| p{0.9cm}<{\centering} p{0.9cm}<{\centering} p{0.9cm}<{\centering}}
        \toprule
        & \multicolumn{2}{c}{} & \multicolumn{3}{c}{\textbf{Full-length Transcript/Video}} 
        & \multicolumn{3}{c}{\textbf{Segmented Transcript/Video}} \\
        \cmidrule(lr){4-9} 
        Model & Size & Note & T & TA & TAC & T & TA & TAC \\
        
        \midrule
        \textbf{Open-source LLMs}  &     &  & &  &  &  &  &  \\
        DeepSeek-R1     &  671B   & MoE & 63.89 & 45.81 & 21.29 & 71.18 & 47.29 & 21.36 \\
        DeepSeek-V3     & 671B & MoE    & 64.91 & 47.10 & \cellcolor{green!20}23.65 & 77.56 & 51.35 & \cellcolor{green!20}28.17 \\
        Llama-3.1     & 8B &            & 57.80 & 40.69 & 19.08 & 71.86 & 43.42 & 21.81 \\
        Llama-3.1     & 70B &           & 63.37 & 45.51 & 19.29 & 76.51 & 50.64 & 23.96 \\   
        Llama-3.1    & 405B &           & 62.83 & 45.82 & 22.67 & 78.60 & 49.50 & 25.38 \\
        Mistral       & 7B &            & 49.94 & 34.78 & 14.13 & 63.73 & 39.23 & 19.83  \\ 
        Mixtral-8x22B & 141B & MoE      & 57.60 & 39.86 & 16.49 & 70.12 & 41.75 & 19.56 \\
        Qwen2.5    & 7B &               & 56.06 & 41.01 & 19.54 & 65.57 & 38.16 & 20.76 \\        
        Qwen2.5 & 72B &                 & 65.65 & \cellcolor{green!20}47.42 & 19.65 & 79.20 & 48.36 & 22.77 \\

        \midrule
        \textbf{Proprietary LLMs}  &     &  & &  &  &  &  &  \\
        
        Claude 3.5 Haiku   & &          & 56.94 & 40.66 & 19.75 & 68.18 & 43.61 & 21.54 \\
        Claude 3.5 Sonnet  & &          & 65.32 & 43.38 & 22.54 & 75.90 & 46.47 & 24.60 \\
        Gemini 1.5 Flash  & &           & 56.90 & 38.20 & 18.97 & 72.09 & 44.83 & 21.31 \\  
        Gemini 2.0 Flash  & &           & 58.36 & 37.47 & 19.20 & 70.05 & 42.11 & 21.43 \\  
        Gemini 1.5 Pro   & &            & 62.78 & 43.25 & 21.26 & 76.56 & 46.38 & 22.87 \\
        Gemini 2.0 Pro  & &             & 60.75 & 44.59 & 18.63 & 74.66 & 47.18 & 24.20 \\        
        GPT-4o        & &               & 57.43 & 37.64 & 15.35 & 76.76 & 45.90 & 24.50 \\
        
        \midrule
        \textbf{Open-source MLLMs}  &     &  & &  &  &  &  &  \\
        
        LLaVa-v1.6-Mistral  & 7B &      & - & - & - & 16.46 & 11.45 & 3.30 \\
        \midrule
        \textbf{Proprietary MLLMs}  &     &  & &  &  &  &  &  \\
        Gemini 1.5 Flash  & &            & 64.91 & 45.06 & 20.66 & \cellcolor{green!20}86.23 & 54.21 & 23.27 \\  
        Gemini 2.0 Flash  &          &   & 64.79 & 40.07 & 19.74 & 82.24 & 48.00 & 23.52 \\ 
        Gemini 1.5 Pro  &           &    & 66.22 & 46.90 & 23.23 & 86.01 & 53.51 & 24.97 \\
        Gemini 2.0 Pro  & &              & \cellcolor{green!20}66.42 & 46.17 & 18.25 & 86.04 & \cellcolor{green!20}54.28 & 25.21 \\ 
        GPT-4o          & &              & 64.95 & 44.53 & 19.60 & 83.47 & 51.15 & 27.86 \\
        \bottomrule
    \end{tabular}
\end{table*}

\section{Experiments}

\subsection{Models}
We evaluate the following LLMs and MLLMs on \datasetname, assessing their performance on textual and multimodal (video) data.

\begin{itemize}
    \item \textbf{Open-source LLMs:} Evaluated on text data. These include DeepSeek-R1 \cite{deepseekai2025deepseekr1incentivizingreasoningcapability}, DeepSeek-V3 \cite{deepseekai2024deepseekv3technicalreport},  Llama-3.1-8B, Llama-3.1-70B, Llama-3.1-405B \cite{grattafiori2024llama3herdmodels}, Mistral-7B-Instruct-v0.2 \cite{jiang2023mistral7b}, Mixtral-8x22B-Instruct-v0.1 \cite{jiang2024mixtralexperts}, Qwen2.5-7B, and Qwen2.5-72B \cite{bai2023qwentechnicalreport}.

    \item \textbf{Proprietary LLMs:} Evaluated on text data. These include Claude 3.5 Haiku, Claude 3.5 Sonnet \cite{anthropic2024claude}, Gemini 1.5 Flash, Gemini 2.0 Flash, Gemini 1.5 Pro, Gemini 2.0 Pro \cite{geminiteam2024geminifamilyhighlycapable}, and GPT-4o \cite{gpt4}.

    \item \textbf{Open-source MLLMs:} We evaluated LLaVA-v1.6-mistral-7B on multimodal (video) data \cite{liu2024improvedbaselinesvisualinstruction}.

    \item \textbf{Proprietary MLLMs:} Evaluated on multimodal (video) data. These include Gemini 1.5 Flash, Gemini 2.0 Flash, Gemini 1.5 Pro, Gemini 2.0 Pro \cite{geminiteam2024geminifamilyhighlycapable} and GPT-4o \cite{gpt4}.
\end{itemize}

Implementation details for each model are in Appendix \ref{app: ImplementationDetails}. Compared to prior work \citep{chandrasegaran2024hourvideo1hourvideolanguageunderstanding}, which evaluates only a limited set of LLMs alongside MLLMs, we provide a more comprehensive benchmarking of both LLMs and MLLMs.

\subsection{Task Suite}
\label{sec:task_suite}

Our sequential task suite spans three tasks: T, TA, and TAC. Task T extracts the recommended ticker, TA includes ticker name and the associated action, and TAC incorporates ticker name, action and conviction score.

This structured progression—\{T, TA, TAC\}—mirrors the logical flow of a financial recommendation. The ticker name serves as the foundation of the recommendation, making it the natural starting point. Once the ticker is identified, the next essential element is the recommended action—what should be done with the identified ticker. Finally, after establishing both the stock ticker name and the action, the conviction score is introduced to evaluate how convincing the finfluencer comes across as. This sequential task suite lays the groundwork for a coherent evaluation framework.

\subsection{Evaluation Framework}
We applied the task suite (T, TA, TAC) to both full-length and segmented videos, allowing us to compare model performance with more noisy full length videos versus focused context from segmented videos. Therefore, we use four different inputs: \textbf{full-length video, segmented video, full-length transcript, and segmented transcript}.

During inference, we used a single prompt to extract three elements—ticker name, action, and conviction score—simultaneously for tasks T, TA, and TAC. During evaluation, we separated the results into T, TA, and TAC to achieve the sequential task suite. We did this because running model inferences separately with three different prompts for tasks T, TA, and TAC would be very computationally expensive, tripling the number of inferences. 

For evaluation, human annotations were used as ground truth, and predicted labels were derived from inference outputs. Performance across all three tasks is evaluated using the F1 score metric. 

\subsection{Results}
Table \ref{tab:benchmarking_results} shows the zero-shot performance of all models in the \datasetname\ benchmark, broken down by task (T, TA, TAC). Our key findings include:

\paragraph{Multimodal Inputs Improve Ticker Extraction (Task T)} The inclusion of multiple modalities significantly enhances ticker extraction performance. In our evaluation, we classify Gemini and GPT models as MLLMs when they receive video as supporting input, leading to better performance on Task T. This improvement is likely due to stock charts in videos displaying ticker names, reducing hallucination errors (e.g., mislabeling Apple's ticker 'AAPL' as 'APPL').

\paragraph{Models Struggle to Identify Investment actions and Conviction (Tasks TA and TAC)} Unlike ticker extraction (Task T), which benefits from visual cues such as stock charts, Tasks TA and TAC require interpreting investment actions and conviction. Models often confuse general market commentary with explicit buy/sell recommendations, leading to frequent misclassifications. Even with multimodal inputs, MLLMs fail to fully capture nuances in speaker intent, tone, and implicit reasoning, highlighting the gap from human-level understanding in financial discourse.

\paragraph{Segmented Inputs Outperform Full-length Inputs} Providing models with more focused, segmented inputs improves performance. Full-length videos often introduce noise—such as sponsorships, promotional content, discussions of unrelated stocks, and general market updates—that dilute relevant information and hinder model effectiveness. By isolating key segments, we reduce these distractions and improve performance.

\paragraph{Significant Gap in Multimodal Understanding} While multimodal inputs improve ticker extraction (Task T), MLLMs struggle with more complex tasks requiring financial reasoning. The proprietary MLLMs perform similarly, while LLaVa-v1.6-Mistral-7B underperforms across all tasks, aligning with \citet{yasunaga2025multimodalrewardbenchholisticevaluation}, which finds that smaller MLLMs lack the domain-specific knowledge needed for complex analysis. Our results further support evidence of U-shaped scaling effects in complex reasoning tasks \cite{zhang2023positivescalingnegationimpacts}. Additionally, open-source MLLMs were difficult to evaluate due to short video context limits ($\leq$ 4 minutes even with a sampling FPS of 0.25).

Although multimodal inputs improve Task T, they provide little benefit for Tasks TA and TAC. The best-performing model for Task T is Gemini 1.5 Flash on segmented video (86.23\%), yet for Task TAC—the most complex task—DeepSeek-V3 (28.17\%) slightly outperforms the top MLLM (27.86\%). This suggests that current LLMs can outperform MLLMs in financial reasoning even without multimodal inputs. With LLMs' stronger performance, lower cost, and reduced computational demands, LLMs remain the more practical choice for some complex tasks (e.g., Task TAC).

\section{Financial Portfolio Analysis}
We analyze portfolio performance using labeled data, as current models struggle with this task. This analysis highlights why extracting key information (Section \ref{sec:dataset}) is crucial for understanding investment outcomes. We evaluate finfluencer recommendations against market indices and examine the role of conviction in stock performance. 

\subsection{Strategies and Market Benchmarks}
Recent trends\footnote{\url{https://www.bis.org/publ/qtrpdf/r_qt2103v.htm}} indicate a growing preference among retail investors for buy-and-hold strategies, particularly in individual stocks. Since 2017, share turnover for S\&P 500 constituents has increased, while turnover for ETFs tracking the index has remained steady, suggesting a shift towards holding individual equities over actively trading index funds. In line with this trend, we evaluated several strategies that a retail investor might adopt after consuming finfluencer content:

\begin{itemize}
    \item \textbf{Buy-and-Hold:} Invest in stocks recommended as "Buy" and hold them for six months \citep{Lim_Rosario_2008b}. 
    
    \item \textbf{Buy-and-Hold Weighted by Conviction:}  
    Allocate capital proportionally to conviction scores, where higher conviction recommendations receive greater allocation. The weight for each asset \( i \) in the portfolio (consisting of all the stocks $j$) is given by \( w_i = \frac{C_i}{\sum_j C_j} \), where \( C_i \) is the conviction score. The allocated capital is then \( \text{Allocation}_i = w_i \times \text{Total Capital} \). 

    \item \textbf{YouTuber Inverse:}  
    Taking a contrarian approach\footnote{\url{https://finance.yahoo.com/quote/SJIM/}. An actively managed ETF that aims to inverse Jim Cramer's stock recommendations.} by doing the opposite of the influencer's advice: if they recommend "Buy," you "Sell," and vice versa. This tests whether opposing the influencer's guidance yields better returns. 
\end{itemize}
 
For benchmarking, we compared these strategies against two popular market indices: 

\begin{itemize}
    \item \textbf{QQQ:}  
    Invest in the NASDAQ-100 index fund (QQQ) and hold it. QQQ tracks major technology companies.
    \item \textbf{S\&P500:}  
    Invest in the S\&P 500 index fund (SPY) and hold it. SPY tracks a basket of 500 large U.S. companies.
\end{itemize}

\subsection{Performance Results}
\begin{figure}[ht]
\centering
\includegraphics[width=\columnwidth]{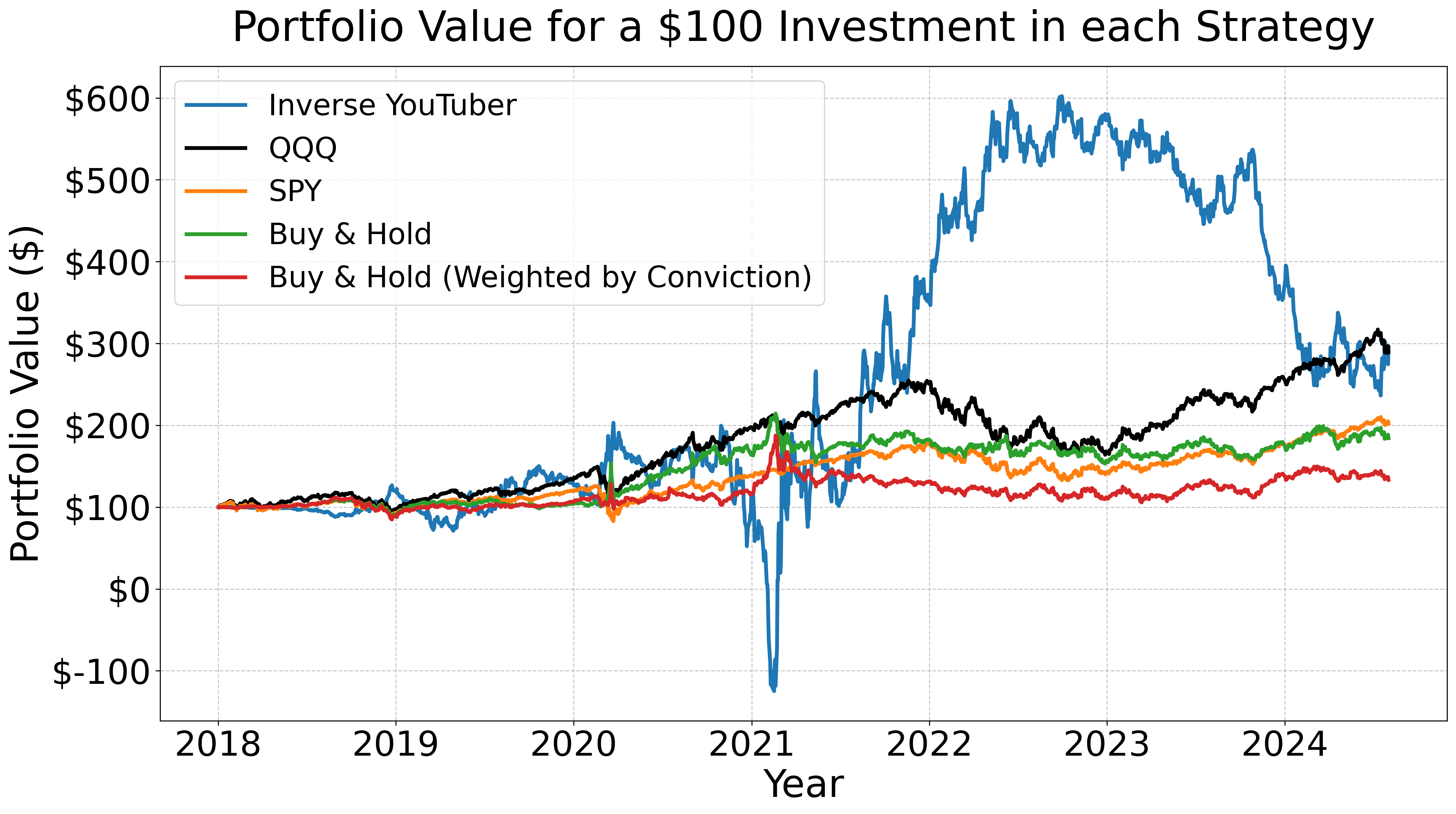}
\caption{Portfolio value on a \$100 investment: The Inverse YouTuber strategy outperforms QQQ and S\&P500, while all other strategies underperform.}
\label{fig:strategies_performance}
\end{figure}

Table~\ref{tab:strategies_performance} and Figure~\ref{fig:strategies_performance} present the performance of various trading strategies from 1st January 2018 to 1st August 2024, each starting with an initial investment of \$100. Performance is assessed using four key metrics: \textit{Sharpe Ratio}\footnote{\url{https://www.investopedia.com/terms/s/sharperatio.asp}}, \textit{Profit and Loss (PnL)}, \textit{Cumulative Return \% (Cumulative)}, and \textit{Annual Return \% (Ann.)}, following calls to adopt finance-specific evaluation criteria that better align with real-world investment outcomes \citep{tatarinov2025languagemodelingfuturefinance}. A detailed financial glossary of terms used is in Appendix \ref{app:financial_gloassary}.

\begin{table}[h]
\caption{Performance of Trading Strategies vs. Market Benchmarks, ordered by \textit{Ann.} (annual return \%). All trades were held for a 6-month period before exit, following \citep{Lim_Rosario_2008b}. }
\label{tab:strategies_performance}
\centering
\resizebox{\columnwidth}{!}{%
\begin{tabular}{lrrrr}
\toprule
\textbf{Strategy/Benchmark} & \textbf{Sharpe} & \textbf{PnL (\$)} & \textbf{Cumulative (\%)} & \textbf{Ann. (\%)} \\
\midrule
Inverse YouTuber    & 0.41  & 195.38  & 195.38  & 17.90  \\
QQQ              & 0.68  & 189.74  & 189.74  & 17.55  \\
SPY              & 0.65  & 102.02  & 102.02  & 11.28  \\
Buy-and-Hold        & 0.46  & 84.29   & 84.29   & 9.74   \\
Buy-and-Hold (Weighted by Conviction)   & 0.30  & 	33.35   & 	33.35  & 4.47   \\
\bottomrule
\end{tabular}%
}
\end{table}

The results reveal that simple buy-and-hold index strategies (QQQ and SPY) consistently outperform finfluencer-based approaches: \textit{Buy \& Hold} and \textit{Buy \& Hold (Weighted)}. Among benchmark indices, QQQ delivers the highest risk-adjusted returns (Sharpe ratio of 0.68), demonstrating the advantage of passive investment in high-growth indices. SPY, while yielding lower returns, maintains stability (second highest Sharpe ratio of 0.65). 

The \textit{Inverse YouTuber} strategy, which takes contrarian positions against finfluencers recommendations, achieves the highest cumulative return but with a lower Sharpe ratio, indicating higher risk. Overall, the findings emphasize that index investing tends to be more effective than finfluencers recommendations, aligning with the broader principle that simplicity often leads to better investment outcomes.

\begin{figure}[ht]
\centering
\includegraphics[width=\columnwidth]{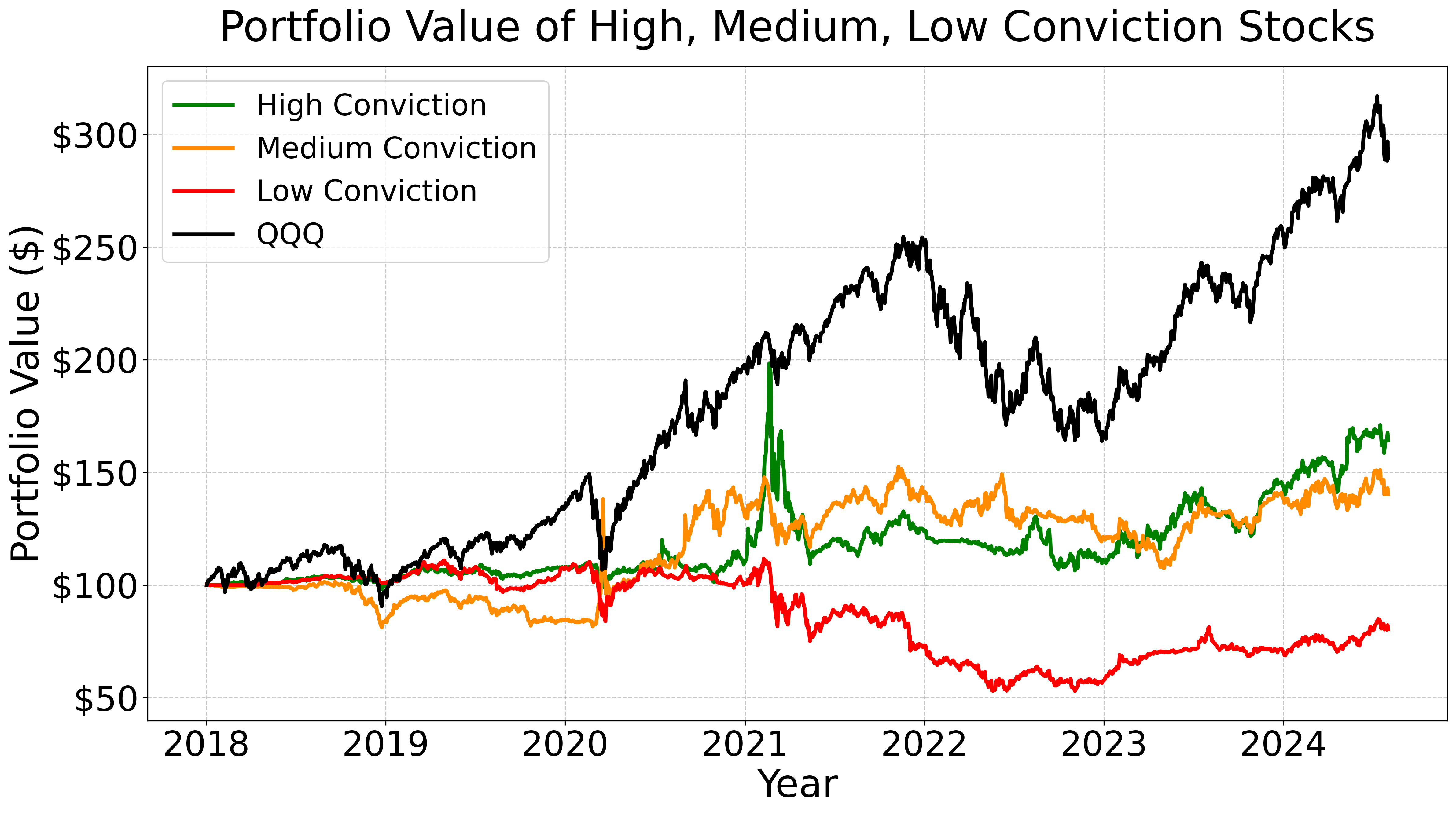}
\caption{Portfolio value of High, Medium, and Low Conviction Stocks (Held for 6 Months). Although high conviction outperforms low and medium conviction, it still lags behind standard benchmarks like QQQ. On a \$100 investment, Total Return is High (64.15\%), Medium (40.41\%), and Low (-19.65\%).}
\label{fig:convictionbuckets}
\end{figure}

\paragraph{Higher Conviction, Better Performance?}
We classified recommendations into low, medium, and high conviction\footnote{Based on Section \ref{sub: expert_human_annotation} expert human annotation. Low conviction = conviction score of 1/3. Medium conviction = conviction score of 2/3. High conviction = conviction score of 3/3.} and assessed their six-month performance. As seen in Figure \ref{fig:convictionbuckets}, while high-conviction picks outperformed low-conviction ones, they still lagged behind the QQQ index fund’s returns. This suggests that while stronger conviction helps avoid the worst outcomes, it does not guarantee to beat a market index. Interestingly, extremely hesitant recommendations performed the worst, reinforcing the idea that a lack of conviction may signal poor investment quality. 

\subsection{Takeaways from Portfolio Analysis}
\begin{itemize} 
    \item \textbf{Best Strategies} – Holding QQQ outperformed all finfluencer based approaches. While the \textit{YouTuber Inverse} strategy (betting against finfluencers) yielded the highest return, it was riskier than market index funds like QQQ. 
    \item \textbf{Most (80\%) finfluencers recommended stocks failed} to beat QQQ, showing that broad-market index funds offer more reliable returns than individual stock picks. For more details see Appendix \ref{ap:quintile}. 
    \item \textbf{Risk of Blindly Following Advice} – Some finfluencers picks perform well, but many significantly underperform, making it a risky strategy. 
    \item \textbf{Higher Conviction, Still Lacking} – High-conviction recommendations (conviction score of 3/3) performed better than low-conviction ones but still failed to beat QQQ.
\end{itemize}

\section{Conclusion and Future Work}
In this paper, we introduce \datasetname, a new benchmark for extracting financial insights from multimodal data. We propose a comprehensive data generation pipeline that includes expert annotation guidelines, transcripts/ASR generation, a structured data collection process, multimodal benchmarking, and financial insight analysis (e.g., portfolio construction). Our evaluation highlights where MLLMs are essential—tasks requiring video and audio modalities for improved performance—versus tasks where text alone suffices (suited for blind LLMs). Our portfolio analysis underscores the critical role of extracting key financial information in understanding investment outcomes. Future work can focus on enhancing MLLMs’ ability to extract empirical financial data, such as price targets. 

\newpage

\bibliographystyle{ACM-Reference-Format}
\balance
\bibliography{ref}


\begin{thebibliography}{62}


\ifx \showCODEN    \undefined \def \showCODEN     #1{\unskip}     \fi
\ifx \showISBNx    \undefined \def \showISBNx     #1{\unskip}     \fi
\ifx \showISBNxiii \undefined \def \showISBNxiii  #1{\unskip}     \fi
\ifx \showISSN     \undefined \def \showISSN      #1{\unskip}     \fi
\ifx \showLCCN     \undefined \def \showLCCN      #1{\unskip}     \fi
\ifx \shownote     \undefined \def \shownote      #1{#1}          \fi
\ifx \showarticletitle \undefined \def \showarticletitle #1{#1}   \fi
\ifx \showURL      \undefined \def \showURL       {\relax}        \fi
\providecommand\bibfield[2]{#2}
\providecommand\bibinfo[2]{#2}
\providecommand\natexlab[1]{#1}
\providecommand\showeprint[2][]{arXiv:#2}

\bibitem[Abelson({[n.\,d.]})]%
        {Abelson}
\bibfield{author}{\bibinfo{person}{Robert~P Abelson}.} \bibinfo{year}{[n.\,d.]}\natexlab{}.
\newblock \bibinfo{title}{Conviction}.
\newblock
\urldef\tempurl%
\url{https://psycnet.apa.org/buy/1988-23399-001}
\showURL{%
\tempurl}


\bibitem[Abu-El-Haija et~al\mbox{.}(2016)]%
        {abuelhaija2016youtube8mlargescalevideoclassification}
\bibfield{author}{\bibinfo{person}{Sami Abu-El-Haija}, \bibinfo{person}{Nisarg Kothari}, \bibinfo{person}{Joonseok Lee}, \bibinfo{person}{Paul Natsev}, \bibinfo{person}{George Toderici}, \bibinfo{person}{Balakrishnan Varadarajan}, {and} \bibinfo{person}{Sudheendra Vijayanarasimhan}.} \bibinfo{year}{2016}\natexlab{}.
\newblock \bibinfo{title}{YouTube-8M: A Large-Scale Video Classification Benchmark}.
\newblock
\showeprint[arxiv]{1609.08675}~[cs.CV]
\urldef\tempurl%
\url{https://arxiv.org/abs/1609.08675}
\showURL{%
\tempurl}


\bibitem[Ang and Lim(2022)]%
        {Ang2022Guided}
\bibfield{author}{\bibinfo{person}{Gary Ang} {and} \bibinfo{person}{Ee-Peng Lim}.} \bibinfo{year}{2022}\natexlab{}.
\newblock \showarticletitle{Guided Attention Multimodal Multitask Financial Forecasting with Inter-Company Relationships and Global and Local News}. In \bibinfo{booktitle}{\emph{Proceedings of the 60th Annual Meeting of the Association for Computational Linguistics (Volume 1: Long Papers)}}, \bibfield{editor}{\bibinfo{person}{Smaranda Muresan}, \bibinfo{person}{Preslav Nakov}, {and} \bibinfo{person}{Aline Villavicencio}} (Eds.). \bibinfo{publisher}{Association for Computational Linguistics}, \bibinfo{address}{Dublin, Ireland}, \bibinfo{pages}{6313--6326}.
\newblock
\href{https://doi.org/10.18653/v1/2022.acl-long.437}{doi:\nolinkurl{10.18653/v1/2022.acl-long.437}}


\bibitem[Anthropic(2024)]%
        {anthropic2024claude}
\bibfield{author}{\bibinfo{person}{Anthropic}.} \bibinfo{year}{2024}\natexlab{}.
\newblock \bibinfo{title}{Introducing the Next Generation of Claude}.
\newblock \bibinfo{howpublished}{\url{https://www.anthropic.com/news/claude-3-family}}.
\newblock
\newblock
\shownote{Accessed: 2024-02-10}.


\bibitem[Araci(2019)]%
        {araci2019finbert}
\bibfield{author}{\bibinfo{person}{Dogu Araci}.} \bibinfo{year}{2019}\natexlab{}.
\newblock \bibinfo{title}{FinBERT: Financial Sentiment Analysis with Pre-trained Language Models}.
\newblock
\showeprint[arxiv]{1908.10063}~[cs.CL]
\urldef\tempurl%
\url{https://arxiv.org/abs/1908.10063}
\showURL{%
\tempurl}


\bibitem[Asness(2024)]%
        {asness2024less}
\bibfield{author}{\bibinfo{person}{Clifford~S Asness}.} \bibinfo{year}{2024}\natexlab{}.
\newblock \showarticletitle{The Less-Efficient Market Hypothesis}.
\newblock \bibinfo{journal}{\emph{Forthcoming in the 50th Anniversary Issue of The Journal of Portfolio Management}} (\bibinfo{year}{2024}).
\newblock


\bibitem[Bagher~Zadeh et~al\mbox{.}(2018)]%
        {bagher-zadeh-etal-2018-multimodal}
\bibfield{author}{\bibinfo{person}{AmirAli Bagher~Zadeh}, \bibinfo{person}{Paul~Pu Liang}, \bibinfo{person}{Soujanya Poria}, \bibinfo{person}{Erik Cambria}, {and} \bibinfo{person}{Louis-Philippe Morency}.} \bibinfo{year}{2018}\natexlab{}.
\newblock \showarticletitle{Multimodal Language Analysis in the Wild: {CMU}-{MOSEI} Dataset and Interpretable Dynamic Fusion Graph}. In \bibinfo{booktitle}{\emph{Proceedings of the 56th Annual Meeting of the Association for Computational Linguistics (Volume 1: Long Papers)}}, \bibfield{editor}{\bibinfo{person}{Iryna Gurevych} {and} \bibinfo{person}{Yusuke Miyao}} (Eds.). \bibinfo{publisher}{Association for Computational Linguistics}, \bibinfo{address}{Melbourne, Australia}, \bibinfo{pages}{2236--2246}.
\newblock
\href{https://doi.org/10.18653/v1/P18-1208}{doi:\nolinkurl{10.18653/v1/P18-1208}}


\bibitem[Chacon et~al\mbox{.}(2023)]%
        {chacon2023will}
\bibfield{author}{\bibinfo{person}{Ryan~G Chacon}, \bibinfo{person}{Thibaut~G Morillon}, {and} \bibinfo{person}{Ruixiang Wang}.} \bibinfo{year}{2023}\natexlab{}.
\newblock \showarticletitle{Will the reddit rebellion take you to the moon? Evidence from WallStreetBets}.
\newblock \bibinfo{journal}{\emph{Financial Markets and Portfolio Management}} \bibinfo{volume}{37}, \bibinfo{number}{1} (\bibinfo{year}{2023}), \bibinfo{pages}{1--25}.
\newblock


\bibitem[Chandrasegaran et~al\mbox{.}(2024)]%
        {chandrasegaran2024hourvideo1hourvideolanguageunderstanding}
\bibfield{author}{\bibinfo{person}{Keshigeyan Chandrasegaran}, \bibinfo{person}{Agrim Gupta}, \bibinfo{person}{Lea~M. Hadzic}, \bibinfo{person}{Taran Kota}, \bibinfo{person}{Jimming He}, \bibinfo{person}{Cristóbal Eyzaguirre}, \bibinfo{person}{Zane Durante}, \bibinfo{person}{Manling Li}, \bibinfo{person}{Jiajun Wu}, {and} \bibinfo{person}{Li Fei-Fei}.} \bibinfo{year}{2024}\natexlab{}.
\newblock \bibinfo{title}{HourVideo: 1-Hour Video-Language Understanding}.
\newblock
\showeprint[arxiv]{2411.04998}~[cs.CV]
\urldef\tempurl%
\url{https://arxiv.org/abs/2411.04998}
\showURL{%
\tempurl}


\bibitem[Das and Singh(2023)]%
        {DasSingh2023}
\bibfield{author}{\bibinfo{person}{Ringki Das} {and} \bibinfo{person}{Thoudam~Doren Singh}.} \bibinfo{year}{2023}\natexlab{}.
\newblock \showarticletitle{Multimodal Sentiment Analysis: A Survey of Methods, Trends, and Challenges}.
\newblock \bibinfo{journal}{\emph{Comput. Surveys}} \bibinfo{volume}{55}, \bibinfo{number}{13s}, Article \bibinfo{articleno}{270} (\bibinfo{date}{jul} \bibinfo{year}{2023}), \bibinfo{numpages}{38}~pages.
\newblock
\showISSN{0360-0300}
\href{https://doi.org/10.1145/3586075}{doi:\nolinkurl{10.1145/3586075}}


\bibitem[Dayoon~Ko(2023)]%
        {ko2023can}
\bibfield{author}{\bibinfo{person}{Gunhee~Kim Dayoon~Ko, Sangho~Lee}.} \bibinfo{year}{2023}\natexlab{}.
\newblock \showarticletitle{Can Language Models Laugh at YouTube Short-form Videos?}. In \bibinfo{booktitle}{\emph{The 2023 Conference on Empirical Methods in Natural Language Processing}}.
\newblock


\bibitem[de~Regt et~al\mbox{.}(2022)]%
        {de2022young}
\bibfield{author}{\bibinfo{person}{Anouk de Regt}, \bibinfo{person}{Zixuan Cheng}, {and} \bibinfo{person}{Rayan Fawaz}.} \bibinfo{year}{2022}\natexlab{}.
\newblock \showarticletitle{Young people under ‘Finfluencer’: The rise of financial influencers on Instagram: An abstract}. In \bibinfo{booktitle}{\emph{Academy of Marketing Science Annual Conference}}. Springer, \bibinfo{pages}{271--272}.
\newblock


\bibitem[{DeepSeek-AI et al.}(2024)]%
        {deepseekai2024deepseekv3technicalreport}
\bibfield{author}{\bibinfo{person}{{DeepSeek-AI et al.}}} \bibinfo{year}{2024}\natexlab{}.
\newblock \bibinfo{title}{DeepSeek-V3 Technical Report}.
\newblock
\showeprint[arxiv]{2412.19437}~[cs.CL]
\urldef\tempurl%
\url{https://arxiv.org/abs/2412.19437}
\showURL{%
\tempurl}


\bibitem[{DeepSeek-AI et al.}(2025)]%
        {deepseekai2025deepseekr1incentivizingreasoningcapability}
\bibfield{author}{\bibinfo{person}{{DeepSeek-AI et al.}}} \bibinfo{year}{2025}\natexlab{}.
\newblock \bibinfo{title}{DeepSeek-R1: Incentivizing Reasoning Capability in LLMs via Reinforcement Learning}.
\newblock
\showeprint[arxiv]{2501.12948}~[cs.CL]
\urldef\tempurl%
\url{https://arxiv.org/abs/2501.12948}
\showURL{%
\tempurl}


\bibitem[{Gemini Team et al.}(2024)]%
        {geminiteam2024geminifamilyhighlycapable}
\bibfield{author}{\bibinfo{person}{{Gemini Team et al.}}} \bibinfo{year}{2024}\natexlab{}.
\newblock \bibinfo{title}{Gemini: A Family of Highly Capable Multimodal Models}.
\newblock
\showeprint[arxiv]{2312.11805}~[cs.CL]
\urldef\tempurl%
\url{https://arxiv.org/abs/2312.11805}
\showURL{%
\tempurl}


\bibitem[Giannini et~al\mbox{.}(2017)]%
        {10.1093/rapstu/rax020}
\bibfield{author}{\bibinfo{person}{Robert Giannini}, \bibinfo{person}{Paul Irvine}, {and} \bibinfo{person}{Tao Shu}.} \bibinfo{year}{2017}\natexlab{}.
\newblock \showarticletitle{{Nonlocal Disadvantage: An Examination of Social Media Sentiment}}.
\newblock \bibinfo{journal}{\emph{The Review of Asset Pricing Studies}} \bibinfo{volume}{8}, \bibinfo{number}{2} (\bibinfo{date}{07} \bibinfo{year}{2017}), \bibinfo{pages}{293--336}.
\newblock
\showISSN{2045-9920}
\href{https://doi.org/10.1093/rapstu/rax020}{doi:\nolinkurl{10.1093/rapstu/rax020}}
\showeprint{https://academic.oup.com/raps/article-pdf/8/2/293/26718371/rax020\_supp.pdf}


\bibitem[Grauman et~al\mbox{.}(2022)]%
        {grauman2022ego4dworld3000hours}
\bibfield{author}{\bibinfo{person}{Kristen Grauman}, \bibinfo{person}{Andrew Westbury}, \bibinfo{person}{Eugene Byrne}, \bibinfo{person}{Zachary Chavis}, \bibinfo{person}{Antonino Furnari}, \bibinfo{person}{Rohit Girdhar}, \bibinfo{person}{Jackson Hamburger}, \bibinfo{person}{Hao Jiang}, \bibinfo{person}{Miao Liu}, \bibinfo{person}{Xingyu Liu}, \bibinfo{person}{Miguel Martin}, \bibinfo{person}{Tushar Nagarajan}, \bibinfo{person}{Ilija Radosavovic}, \bibinfo{person}{Santhosh~Kumar Ramakrishnan}, \bibinfo{person}{Fiona Ryan}, \bibinfo{person}{Jayant Sharma}, \bibinfo{person}{Michael Wray}, \bibinfo{person}{Mengmeng Xu}, \bibinfo{person}{Eric~Zhongcong Xu}, \bibinfo{person}{Chen Zhao}, \bibinfo{person}{Siddhant Bansal}, \bibinfo{person}{Dhruv Batra}, \bibinfo{person}{Vincent Cartillier}, \bibinfo{person}{Sean Crane}, \bibinfo{person}{Tien Do}, \bibinfo{person}{Morrie Doulaty}, \bibinfo{person}{Akshay Erapalli}, \bibinfo{person}{Christoph Feichtenhofer}, \bibinfo{person}{Adriano Fragomeni},
  \bibinfo{person}{Qichen Fu}, \bibinfo{person}{Abrham Gebreselasie}, \bibinfo{person}{Cristina Gonzalez}, \bibinfo{person}{James Hillis}, \bibinfo{person}{Xuhua Huang}, \bibinfo{person}{Yifei Huang}, \bibinfo{person}{Wenqi Jia}, \bibinfo{person}{Weslie Khoo}, \bibinfo{person}{Jachym Kolar}, \bibinfo{person}{Satwik Kottur}, \bibinfo{person}{Anurag Kumar}, \bibinfo{person}{Federico Landini}, \bibinfo{person}{Chao Li}, \bibinfo{person}{Yanghao Li}, \bibinfo{person}{Zhenqiang Li}, \bibinfo{person}{Karttikeya Mangalam}, \bibinfo{person}{Raghava Modhugu}, \bibinfo{person}{Jonathan Munro}, \bibinfo{person}{Tullie Murrell}, \bibinfo{person}{Takumi Nishiyasu}, \bibinfo{person}{Will Price}, \bibinfo{person}{Paola~Ruiz Puentes}, \bibinfo{person}{Merey Ramazanova}, \bibinfo{person}{Leda Sari}, \bibinfo{person}{Kiran Somasundaram}, \bibinfo{person}{Audrey Southerland}, \bibinfo{person}{Yusuke Sugano}, \bibinfo{person}{Ruijie Tao}, \bibinfo{person}{Minh Vo}, \bibinfo{person}{Yuchen Wang}, \bibinfo{person}{Xindi Wu},
  \bibinfo{person}{Takuma Yagi}, \bibinfo{person}{Ziwei Zhao}, \bibinfo{person}{Yunyi Zhu}, \bibinfo{person}{Pablo Arbelaez}, \bibinfo{person}{David Crandall}, \bibinfo{person}{Dima Damen}, \bibinfo{person}{Giovanni~Maria Farinella}, \bibinfo{person}{Christian Fuegen}, \bibinfo{person}{Bernard Ghanem}, \bibinfo{person}{Vamsi~Krishna Ithapu}, \bibinfo{person}{C.~V. Jawahar}, \bibinfo{person}{Hanbyul Joo}, \bibinfo{person}{Kris Kitani}, \bibinfo{person}{Haizhou Li}, \bibinfo{person}{Richard Newcombe}, \bibinfo{person}{Aude Oliva}, \bibinfo{person}{Hyun~Soo Park}, \bibinfo{person}{James~M. Rehg}, \bibinfo{person}{Yoichi Sato}, \bibinfo{person}{Jianbo Shi}, \bibinfo{person}{Mike~Zheng Shou}, \bibinfo{person}{Antonio Torralba}, \bibinfo{person}{Lorenzo Torresani}, \bibinfo{person}{Mingfei Yan}, {and} \bibinfo{person}{Jitendra Malik}.} \bibinfo{year}{2022}\natexlab{}.
\newblock \bibinfo{title}{Ego4D: Around the World in 3,000 Hours of Egocentric Video}.
\newblock
\showeprint[arxiv]{2110.07058}~[cs.CV]
\urldef\tempurl%
\url{https://arxiv.org/abs/2110.07058}
\showURL{%
\tempurl}


\bibitem[Groß-Klußmann et~al\mbox{.}(2019)]%
        {GrossKlussmann2019}
\bibfield{author}{\bibinfo{person}{Axel Groß-Klußmann}, \bibinfo{person}{Stephan K{\"o}nig}, {and} \bibinfo{person}{Markus Ebner}.} \bibinfo{year}{2019}\natexlab{}.
\newblock \showarticletitle{Buzzwords Build Momentum: Global Financial Twitter Sentiment and the Aggregate Stock Market}.
\newblock \bibinfo{journal}{\emph{Expert Systems with Applications}}  \bibinfo{volume}{136} (\bibinfo{year}{2019}), \bibinfo{pages}{171--186}.
\newblock
\showISSN{0957-4174}
\href{https://doi.org/10.1016/j.eswa.2019.06.027}{doi:\nolinkurl{10.1016/j.eswa.2019.06.027}}
\newblock
\shownote{Preprint available at SSRN: \url{https://ssrn.com/abstract=3412908} or \url{http://dx.doi.org/10.2139/ssrn.3412908}}.


\bibitem[Guan(2023)]%
        {Guan_2023}
\bibfield{author}{\bibinfo{person}{Sue Guan}.} \bibinfo{year}{2023}\natexlab{}.
\newblock \showarticletitle{The rise of the Finfluencer}.
\newblock \bibinfo{journal}{\emph{SSRN Electronic Journal}} (\bibinfo{year}{2023}).
\newblock
\href{https://doi.org/10.2139/ssrn.4400042}{doi:\nolinkurl{10.2139/ssrn.4400042}}


\bibitem[Heilbron et~al\mbox{.}(2015)]%
        {caba2015activitynet}
\bibfield{author}{\bibinfo{person}{Fabian~Caba Heilbron}, \bibinfo{person}{Victor Escorcia}, \bibinfo{person}{Bernard Ghanem}, {and} \bibinfo{person}{Juan~Carlos Niebles}.} \bibinfo{year}{2015}\natexlab{}.
\newblock \showarticletitle{ActivityNet: A large-scale video benchmark for human activity understanding}. In \bibinfo{booktitle}{\emph{2015 IEEE Conference on Computer Vision and Pattern Recognition (CVPR)}}. \bibinfo{pages}{961--970}.
\newblock
\href{https://doi.org/10.1109/CVPR.2015.7298698}{doi:\nolinkurl{10.1109/CVPR.2015.7298698}}


\bibitem[Hull and Qi(2024)]%
        {Hull_Qi_2024b}
\bibfield{author}{\bibinfo{person}{Isaiah Hull} {and} \bibinfo{person}{Yingjie Qi}.} \bibinfo{year}{2024}\natexlab{}.
\newblock \bibinfo{journal}{\emph{The impact of Finfluencers on Retail Investment}} (\bibinfo{year}{2024}).
\newblock
\href{https://doi.org/10.2139/ssrn.4922031}{doi:\nolinkurl{10.2139/ssrn.4922031}}


\bibitem[Jiang et~al\mbox{.}(2023)]%
        {jiang2023mistral7b}
\bibfield{author}{\bibinfo{person}{Albert~Q. Jiang}, \bibinfo{person}{Alexandre Sablayrolles}, \bibinfo{person}{Arthur Mensch}, \bibinfo{person}{Chris Bamford}, \bibinfo{person}{Devendra~Singh Chaplot}, \bibinfo{person}{Diego de~las Casas}, \bibinfo{person}{Florian Bressand}, \bibinfo{person}{Gianna Lengyel}, \bibinfo{person}{Guillaume Lample}, \bibinfo{person}{Lucile Saulnier}, \bibinfo{person}{Lélio~Renard Lavaud}, \bibinfo{person}{Marie-Anne Lachaux}, \bibinfo{person}{Pierre Stock}, \bibinfo{person}{Teven~Le Scao}, \bibinfo{person}{Thibaut Lavril}, \bibinfo{person}{Thomas Wang}, \bibinfo{person}{Timothée Lacroix}, {and} \bibinfo{person}{William~El Sayed}.} \bibinfo{year}{2023}\natexlab{}.
\newblock \bibinfo{title}{Mistral 7B}.
\newblock
\showeprint[arxiv]{2310.06825}~[cs.CL]
\urldef\tempurl%
\url{https://arxiv.org/abs/2310.06825}
\showURL{%
\tempurl}


\bibitem[Jiang et~al\mbox{.}(2024)]%
        {jiang2024mixtralexperts}
\bibfield{author}{\bibinfo{person}{Albert~Q. Jiang}, \bibinfo{person}{Alexandre Sablayrolles}, \bibinfo{person}{Antoine Roux}, \bibinfo{person}{Arthur Mensch}, \bibinfo{person}{Blanche Savary}, \bibinfo{person}{Chris Bamford}, \bibinfo{person}{Devendra~Singh Chaplot}, \bibinfo{person}{Diego de~las Casas}, \bibinfo{person}{Emma~Bou Hanna}, \bibinfo{person}{Florian Bressand}, \bibinfo{person}{Gianna Lengyel}, \bibinfo{person}{Guillaume Bour}, \bibinfo{person}{Guillaume Lample}, \bibinfo{person}{Lélio~Renard Lavaud}, \bibinfo{person}{Lucile Saulnier}, \bibinfo{person}{Marie-Anne Lachaux}, \bibinfo{person}{Pierre Stock}, \bibinfo{person}{Sandeep Subramanian}, \bibinfo{person}{Sophia Yang}, \bibinfo{person}{Szymon Antoniak}, \bibinfo{person}{Teven~Le Scao}, \bibinfo{person}{Théophile Gervet}, \bibinfo{person}{Thibaut Lavril}, \bibinfo{person}{Thomas Wang}, \bibinfo{person}{Timothée Lacroix}, {and} \bibinfo{person}{William~El Sayed}.} \bibinfo{year}{2024}\natexlab{}.
\newblock \bibinfo{title}{Mixtral of Experts}.
\newblock
\showeprint[arxiv]{2401.04088}~[cs.LG]
\urldef\tempurl%
\url{https://arxiv.org/abs/2401.04088}
\showURL{%
\tempurl}


\bibitem[Jin et~al\mbox{.}(2024)]%
        {jin2024rjuameddqamultimodalbenchmarkmedical}
\bibfield{author}{\bibinfo{person}{Congyun Jin}, \bibinfo{person}{Ming Zhang}, \bibinfo{person}{Xiaowei Ma}, \bibinfo{person}{Li Yujiao}, \bibinfo{person}{Yingbo Wang}, \bibinfo{person}{Yabo Jia}, \bibinfo{person}{Yuliang Du}, \bibinfo{person}{Tao Sun}, \bibinfo{person}{Haowen Wang}, \bibinfo{person}{Cong Fan}, \bibinfo{person}{Jinjie Gu}, \bibinfo{person}{Chenfei Chi}, \bibinfo{person}{Xiangguo Lv}, \bibinfo{person}{Fangzhou Li}, \bibinfo{person}{Wei Xue}, {and} \bibinfo{person}{Yiran Huang}.} \bibinfo{year}{2024}\natexlab{}.
\newblock \bibinfo{title}{RJUA-MedDQA: A Multimodal Benchmark for Medical Document Question Answering and Clinical Reasoning}.
\newblock
\showeprint[arxiv]{2402.14840}~[cs.CL]
\urldef\tempurl%
\url{https://arxiv.org/abs/2402.14840}
\showURL{%
\tempurl}


\bibitem[{Jinze Bai et al.}(2023)]%
        {bai2023qwentechnicalreport}
\bibfield{author}{\bibinfo{person}{{Jinze Bai et al.}}} \bibinfo{year}{2023}\natexlab{}.
\newblock \bibinfo{title}{Qwen Technical Report}.
\newblock
\showeprint[arxiv]{2309.16609}~[cs.CL]
\urldef\tempurl%
\url{https://arxiv.org/abs/2309.16609}
\showURL{%
\tempurl}


\bibitem[Kakhbod et~al\mbox{.}(2023)]%
        {kakhbod2023finfluencers}
\bibfield{author}{\bibinfo{person}{Ali Kakhbod}, \bibinfo{person}{Seyed~Mohammad Kazempour}, \bibinfo{person}{Dmitry Livdan}, {and} \bibinfo{person}{Norman Schuerhoff}.} \bibinfo{year}{2023}\natexlab{}.
\newblock \bibinfo{booktitle}{\emph{Finfluencers}}.
\newblock \bibinfo{type}{Research Paper} 23-30. \bibinfo{institution}{Swiss Finance Institute}.
\newblock
\newblock
\shownote{Available at SSRN: \url{https://ssrn.com/abstract=4428232} or \url{http://dx.doi.org/10.2139/ssrn.4428232}}.


\bibitem[Krishna et~al\mbox{.}(2017)]%
        {krishna2017densecaptioningeventsvideos}
\bibfield{author}{\bibinfo{person}{Ranjay Krishna}, \bibinfo{person}{Kenji Hata}, \bibinfo{person}{Frederic Ren}, \bibinfo{person}{Li Fei-Fei}, {and} \bibinfo{person}{Juan~Carlos Niebles}.} \bibinfo{year}{2017}\natexlab{}.
\newblock \bibinfo{title}{Dense-Captioning Events in Videos}.
\newblock
\showeprint[arxiv]{1705.00754}~[cs.CV]
\urldef\tempurl%
\url{https://arxiv.org/abs/1705.00754}
\showURL{%
\tempurl}


\bibitem[Lee and Yoo(2020)]%
        {Lee2020}
\bibfield{author}{\bibinfo{person}{Sang~Il Lee} {and} \bibinfo{person}{Seong~Joon Yoo}.} \bibinfo{year}{2020}\natexlab{}.
\newblock \showarticletitle{Multimodal Deep Learning for Finance: Integrating and Forecasting International Stock Markets}.
\newblock \bibinfo{journal}{\emph{The Journal of Supercomputing}} \bibinfo{volume}{76}, \bibinfo{number}{10} (\bibinfo{year}{2020}), \bibinfo{pages}{8294--8312}.
\newblock
\href{https://doi.org/10.1007/s11227-019-03101-3}{doi:\nolinkurl{10.1007/s11227-019-03101-3}}


\bibitem[Liang et~al\mbox{.}(2024)]%
        {liang2024enhancingfinancialmarketpredictions}
\bibfield{author}{\bibinfo{person}{Wenhao Liang}, \bibinfo{person}{Zhengyang Li}, {and} \bibinfo{person}{Weitong Chen}.} \bibinfo{year}{2024}\natexlab{}.
\newblock \bibinfo{title}{Enhancing Financial Market Predictions: Causality-Driven Feature Selection}.
\newblock
\showeprint[arxiv]{2408.01005}~[cs.LG]
\urldef\tempurl%
\url{https://arxiv.org/abs/2408.01005}
\showURL{%
\tempurl}


\bibitem[Lim and Rosario(2008)]%
        {Lim_Rosario_2008b}
\bibfield{author}{\bibinfo{person}{Bryan Lim} {and} \bibinfo{person}{Joao Rosario}.} \bibinfo{year}{2008}\natexlab{}.
\newblock \showarticletitle{The performance and impact of stock picks mentioned on “mad money”}.
\newblock \bibinfo{journal}{\emph{SSRN Electronic Journal}} (\bibinfo{year}{2008}).
\newblock
\href{https://doi.org/10.2139/ssrn.1017353}{doi:\nolinkurl{10.2139/ssrn.1017353}}


\bibitem[Liu et~al\mbox{.}(2024b)]%
        {liu2024improvedbaselinesvisualinstruction}
\bibfield{author}{\bibinfo{person}{Haotian Liu}, \bibinfo{person}{Chunyuan Li}, \bibinfo{person}{Yuheng Li}, {and} \bibinfo{person}{Yong~Jae Lee}.} \bibinfo{year}{2024}\natexlab{b}.
\newblock \bibinfo{title}{Improved Baselines with Visual Instruction Tuning}.
\newblock
\showeprint[arxiv]{2310.03744}~[cs.CV]
\urldef\tempurl%
\url{https://arxiv.org/abs/2310.03744}
\showURL{%
\tempurl}


\bibitem[Liu et~al\mbox{.}(2024c)]%
        {liu2024devandensevideoannotation}
\bibfield{author}{\bibinfo{person}{Tingkai Liu}, \bibinfo{person}{Yunzhe Tao}, \bibinfo{person}{Haogeng Liu}, \bibinfo{person}{Qihang Fan}, \bibinfo{person}{Ding Zhou}, \bibinfo{person}{Huaibo Huang}, \bibinfo{person}{Ran He}, {and} \bibinfo{person}{Hongxia Yang}.} \bibinfo{year}{2024}\natexlab{c}.
\newblock \bibinfo{title}{DeVAn: Dense Video Annotation for Video-Language Models}.
\newblock
\showeprint[arxiv]{2310.05060}~[cs.CV]
\urldef\tempurl%
\url{https://arxiv.org/abs/2310.05060}
\showURL{%
\tempurl}


\bibitem[Liu et~al\mbox{.}(2024a)]%
        {liu2024mmbenchmultimodalmodelallaround}
\bibfield{author}{\bibinfo{person}{Yuan Liu}, \bibinfo{person}{Haodong Duan}, \bibinfo{person}{Yuanhan Zhang}, \bibinfo{person}{Bo Li}, \bibinfo{person}{Songyang Zhang}, \bibinfo{person}{Wangbo Zhao}, \bibinfo{person}{Yike Yuan}, \bibinfo{person}{Jiaqi Wang}, \bibinfo{person}{Conghui He}, \bibinfo{person}{Ziwei Liu}, \bibinfo{person}{Kai Chen}, {and} \bibinfo{person}{Dahua Lin}.} \bibinfo{year}{2024}\natexlab{a}.
\newblock \bibinfo{title}{MMBench: Is Your Multi-modal Model an All-around Player?}
\newblock
\showeprint[arxiv]{2307.06281}~[cs.CV]
\urldef\tempurl%
\url{https://arxiv.org/abs/2307.06281}
\showURL{%
\tempurl}


\bibitem[Loughran and McDonald(2011)]%
        {loughran2011liability}
\bibfield{author}{\bibinfo{person}{Tim Loughran} {and} \bibinfo{person}{Bill McDonald}.} \bibinfo{year}{2011}\natexlab{}.
\newblock \showarticletitle{When is a liability not a liability? Textual analysis, dictionaries, and 10-Ks}.
\newblock \bibinfo{journal}{\emph{The Journal of finance}} \bibinfo{volume}{66}, \bibinfo{number}{1} (\bibinfo{year}{2011}), \bibinfo{pages}{35--65}.
\newblock


\bibitem[Lu et~al\mbox{.}(2023)]%
        {lu2023bbt}
\bibfield{author}{\bibinfo{person}{Dakuan Lu}, \bibinfo{person}{Hengkui Wu}, \bibinfo{person}{Jiaqing Liang}, \bibinfo{person}{Yipei Xu}, \bibinfo{person}{Qianyu He}, \bibinfo{person}{Yipeng Geng}, \bibinfo{person}{Mengkun Han}, \bibinfo{person}{Yingsi Xin}, {and} \bibinfo{person}{Yanghua Xiao}.} \bibinfo{year}{2023}\natexlab{}.
\newblock \showarticletitle{Bbt-fin: Comprehensive construction of chinese financial domain pre-trained language model, corpus and benchmark}.
\newblock \bibinfo{journal}{\emph{arXiv preprint arXiv:2302.09432}} (\bibinfo{year}{2023}).
\newblock


\bibitem[Maia et~al\mbox{.}(2018)]%
        {FiQA}
\bibfield{author}{\bibinfo{person}{Macedo Maia}, \bibinfo{person}{Siegfried Handschuh}, \bibinfo{person}{Andr\'{e} Freitas}, \bibinfo{person}{Brian Davis}, \bibinfo{person}{Ross McDermott}, \bibinfo{person}{Manel Zarrouk}, {and} \bibinfo{person}{Alexandra Balahur}.} \bibinfo{year}{2018}\natexlab{}.
\newblock \showarticletitle{WWW'18 Open Challenge: Financial Opinion Mining and Question Answering}. In \bibinfo{booktitle}{\emph{Companion Proceedings of the The Web Conference 2018}} (Lyon, France) \emph{(\bibinfo{series}{WWW '18})}. \bibinfo{publisher}{International World Wide Web Conferences Steering Committee}, \bibinfo{address}{Republic and Canton of Geneva, CHE}, \bibinfo{pages}{1941–1942}.
\newblock
\showISBNx{9781450356404}
\href{https://doi.org/10.1145/3184558.3192301}{doi:\nolinkurl{10.1145/3184558.3192301}}


\bibitem[Malo et~al\mbox{.}(2013)]%
        {malo2013gooddebtbaddebt}
\bibfield{author}{\bibinfo{person}{Pekka Malo}, \bibinfo{person}{Ankur Sinha}, \bibinfo{person}{Pyry Takala}, \bibinfo{person}{Pekka Korhonen}, {and} \bibinfo{person}{Jyrki Wallenius}.} \bibinfo{year}{2013}\natexlab{}.
\newblock \bibinfo{title}{Good Debt or Bad Debt: Detecting Semantic Orientations in Economic Texts}.
\newblock
\showeprint[arxiv]{1307.5336}~[cs.CL]
\urldef\tempurl%
\url{https://arxiv.org/abs/1307.5336}
\showURL{%
\tempurl}


\bibitem[{Meta et al.}(2024)]%
        {grattafiori2024llama3herdmodels}
\bibfield{author}{\bibinfo{person}{{Meta et al.}}} \bibinfo{year}{2024}\natexlab{}.
\newblock \bibinfo{title}{The Llama 3 Herd of Models}.
\newblock
\showeprint[arxiv]{2407.21783}~[cs.AI]
\urldef\tempurl%
\url{https://arxiv.org/abs/2407.21783}
\showURL{%
\tempurl}


\bibitem[Miech et~al\mbox{.}(2019)]%
        {miech19howto100m}
\bibfield{author}{\bibinfo{person}{Antoine Miech}, \bibinfo{person}{Dimitri Zhukov}, \bibinfo{person}{Jean-Baptiste Alayrac}, \bibinfo{person}{Makarand Tapaswi}, \bibinfo{person}{Ivan Laptev}, {and} \bibinfo{person}{Josef Sivic}.} \bibinfo{year}{2019}\natexlab{}.
\newblock \showarticletitle{How{T}o100{M}: {L}earning a {T}ext-{V}ideo {E}mbedding by {W}atching {H}undred {M}illion {N}arrated {V}ideo {C}lips}. In \bibinfo{booktitle}{\emph{ICCV}}.
\newblock


\bibitem[Ning et~al\mbox{.}(2023)]%
        {ning2023videobenchcomprehensivebenchmarktoolkit}
\bibfield{author}{\bibinfo{person}{Munan Ning}, \bibinfo{person}{Bin Zhu}, \bibinfo{person}{Yujia Xie}, \bibinfo{person}{Bin Lin}, \bibinfo{person}{Jiaxi Cui}, \bibinfo{person}{Lu Yuan}, \bibinfo{person}{Dongdong Chen}, {and} \bibinfo{person}{Li Yuan}.} \bibinfo{year}{2023}\natexlab{}.
\newblock \bibinfo{title}{Video-Bench: A Comprehensive Benchmark and Toolkit for Evaluating Video-based Large Language Models}.
\newblock
\showeprint[arxiv]{2311.16103}~[cs.CV]
\urldef\tempurl%
\url{https://arxiv.org/abs/2311.16103}
\showURL{%
\tempurl}


\bibitem[{OpenAI}(023a)]%
        {gpt4}
\bibfield{author}{\bibinfo{person}{{OpenAI}}.} \bibinfo{year}{2023a}\natexlab{}.
\newblock \bibinfo{booktitle}{\emph{GPT-4 Technical Report}}.
\newblock \bibinfo{type}{Technical Report}. \bibinfo{institution}{OpenAI}.
\newblock
\newblock
\shownote{Available at \url{https://doi.org/10.48550/arXiv.2303.08774}}.


\bibitem[Pardawala et~al\mbox{.}(2024)]%
        {pardawala2024subjectiveqameasuringsubjectivityearnings}
\bibfield{author}{\bibinfo{person}{Huzaifa Pardawala}, \bibinfo{person}{Siddhant Sukhani}, \bibinfo{person}{Agam Shah}, \bibinfo{person}{Veer Kejriwal}, \bibinfo{person}{Abhishek Pillai}, \bibinfo{person}{Rohan Bhasin}, \bibinfo{person}{Andrew DiBiasio}, \bibinfo{person}{Tarun Mandapati}, \bibinfo{person}{Dhruv Adha}, {and} \bibinfo{person}{Sudheer Chava}.} \bibinfo{year}{2024}\natexlab{}.
\newblock \bibinfo{title}{SubjECTive-QA: Measuring Subjectivity in Earnings Call Transcripts' QA Through Six-Dimensional Feature Analysis}.
\newblock
\showeprint[arxiv]{2410.20651}~[cs.CL]
\urldef\tempurl%
\url{https://arxiv.org/abs/2410.20651}
\showURL{%
\tempurl}


\bibitem[Pei et~al\mbox{.}(2022)]%
        {pei-etal-2022-tweetfinsent}
\bibfield{author}{\bibinfo{person}{Yulong Pei}, \bibinfo{person}{Amarachi Mbakwe}, \bibinfo{person}{Akshat Gupta}, \bibinfo{person}{Salwa Alamir}, \bibinfo{person}{Hanxuan Lin}, \bibinfo{person}{Xiaomo Liu}, {and} \bibinfo{person}{Sameena Shah}.} \bibinfo{year}{2022}\natexlab{}.
\newblock \showarticletitle{{T}weet{F}in{S}ent: A Dataset of Stock Sentiments on {T}witter}. In \bibinfo{booktitle}{\emph{Proceedings of the Fourth Workshop on Financial Technology and Natural Language Processing (FinNLP)}}, \bibfield{editor}{\bibinfo{person}{Chung-Chi Chen}, \bibinfo{person}{Hen-Hsen Huang}, \bibinfo{person}{Hiroya Takamura}, {and} \bibinfo{person}{Hsin-Hsi Chen}} (Eds.). \bibinfo{publisher}{Association for Computational Linguistics}, \bibinfo{address}{Abu Dhabi, United Arab Emirates (Hybrid)}, \bibinfo{pages}{37--47}.
\newblock
\href{https://doi.org/10.18653/v1/2022.finnlp-1.5}{doi:\nolinkurl{10.18653/v1/2022.finnlp-1.5}}


\bibitem[Radford et~al\mbox{.}(2023)]%
        {pmlr-v202-radford23a}
\bibfield{author}{\bibinfo{person}{Alec Radford}, \bibinfo{person}{Jong~Wook Kim}, \bibinfo{person}{Tao Xu}, \bibinfo{person}{Greg Brockman}, \bibinfo{person}{Christine Mcleavey}, {and} \bibinfo{person}{Ilya Sutskever}.} \bibinfo{year}{2023}\natexlab{}.
\newblock \showarticletitle{Robust Speech Recognition via Large-Scale Weak Supervision}. In \bibinfo{booktitle}{\emph{Proceedings of the 40th International Conference on Machine Learning}} \emph{(\bibinfo{series}{Proceedings of Machine Learning Research}, Vol.~\bibinfo{volume}{202})}, \bibfield{editor}{\bibinfo{person}{Andreas Krause}, \bibinfo{person}{Emma Brunskill}, \bibinfo{person}{Kyunghyun Cho}, \bibinfo{person}{Barbara Engelhardt}, \bibinfo{person}{Sivan Sabato}, {and} \bibinfo{person}{Jonathan Scarlett}} (Eds.). \bibinfo{publisher}{PMLR}, \bibinfo{pages}{28492--28518}.
\newblock
\urldef\tempurl%
\url{https://proceedings.mlr.press/v202/radford23a.html}
\showURL{%
\tempurl}


\bibitem[Rai et~al\mbox{.}(2024)]%
        {Rai2024NPTEL}
\bibfield{author}{\bibinfo{person}{A.K. Rai}, \bibinfo{person}{S.D. Jaiswal}, {and} \bibinfo{person}{A. Mukherjee}.} \bibinfo{year}{2024}\natexlab{}.
\newblock \showarticletitle{A Deep Dive into the Disparity of Word Error Rates across Thousands of NPTEL MOOC Videos}.
\newblock \bibinfo{journal}{\emph{Proceedings of the International AAAI Conference on Web and Social Media}} \bibinfo{volume}{18}, \bibinfo{number}{1} (\bibinfo{date}{May} \bibinfo{year}{2024}), \bibinfo{pages}{1302--1314}.
\newblock
\href{https://doi.org/10.1609/icwsm.v18i1.31390}{doi:\nolinkurl{10.1609/icwsm.v18i1.31390}}


\bibitem[Sawhney et~al\mbox{.}(2020)]%
        {sawhney-etal-2020-voltage}
\bibfield{author}{\bibinfo{person}{Ramit Sawhney}, \bibinfo{person}{Piyush Khanna}, \bibinfo{person}{Arshiya Aggarwal}, \bibinfo{person}{Taru Jain}, \bibinfo{person}{Puneet Mathur}, {and} \bibinfo{person}{Rajiv~Ratn Shah}.} \bibinfo{year}{2020}\natexlab{}.
\newblock \showarticletitle{{V}ol{TAGE}: Volatility Forecasting via Text Audio Fusion with Graph Convolution Networks for Earnings Calls}. In \bibinfo{booktitle}{\emph{Proceedings of the 2020 Conference on Empirical Methods in Natural Language Processing (EMNLP)}}, \bibfield{editor}{\bibinfo{person}{Bonnie Webber}, \bibinfo{person}{Trevor Cohn}, \bibinfo{person}{Yulan He}, {and} \bibinfo{person}{Yang Liu}} (Eds.). \bibinfo{publisher}{Association for Computational Linguistics}, \bibinfo{address}{Online}, \bibinfo{pages}{8001--8013}.
\newblock
\href{https://doi.org/10.18653/v1/2020.emnlp-main.643}{doi:\nolinkurl{10.18653/v1/2020.emnlp-main.643}}


\bibitem[Shah et~al\mbox{.}(2023)]%
        {shah-etal-2023-trillion}
\bibfield{author}{\bibinfo{person}{Agam Shah}, \bibinfo{person}{Suvan Paturi}, {and} \bibinfo{person}{Sudheer Chava}.} \bibinfo{year}{2023}\natexlab{}.
\newblock \showarticletitle{Trillion Dollar Words: A New Financial Dataset, Task {\&} Market Analysis}. In \bibinfo{booktitle}{\emph{Proceedings of the 61st Annual Meeting of the Association for Computational Linguistics (Volume 1: Long Papers)}}, \bibfield{editor}{\bibinfo{person}{Anna Rogers}, \bibinfo{person}{Jordan Boyd-Graber}, {and} \bibinfo{person}{Naoaki Okazaki}} (Eds.). \bibinfo{publisher}{Association for Computational Linguistics}, \bibinfo{address}{Toronto, Canada}, \bibinfo{pages}{6664--6679}.
\newblock
\href{https://doi.org/10.18653/v1/2023.acl-long.368}{doi:\nolinkurl{10.18653/v1/2023.acl-long.368}}


\bibitem[Shah et~al\mbox{.}(2025)]%
        {shah2025wordsuniteworldunified}
\bibfield{author}{\bibinfo{person}{Agam Shah}, \bibinfo{person}{Siddhant Sukhani}, \bibinfo{person}{Huzaifa Pardawala}, \bibinfo{person}{Saketh Budideti}, \bibinfo{person}{Riya Bhadani}, \bibinfo{person}{Rudra Gopal}, \bibinfo{person}{Siddhartha Somani}, \bibinfo{person}{Michael Galarnyk}, \bibinfo{person}{Soungmin Lee}, \bibinfo{person}{Arnav Hiray}, \bibinfo{person}{Akshar Ravichandran}, \bibinfo{person}{Eric Kim}, \bibinfo{person}{Pranav Aluru}, \bibinfo{person}{Joshua Zhang}, \bibinfo{person}{Sebastian Jaskowski}, \bibinfo{person}{Veer Guda}, \bibinfo{person}{Meghaj Tarte}, \bibinfo{person}{Liqin Ye}, \bibinfo{person}{Spencer Gosden}, \bibinfo{person}{Rutwik Routu}, \bibinfo{person}{Rachel Yuh}, \bibinfo{person}{Sloka Chava}, \bibinfo{person}{Sahasra Chava}, \bibinfo{person}{Dylan~Patrick Kelly}, \bibinfo{person}{Aiden Chiang}, \bibinfo{person}{Harsit Mittal}, {and} \bibinfo{person}{Sudheer Chava}.} \bibinfo{year}{2025}\natexlab{}.
\newblock \bibinfo{title}{Words That Unite The World: A Unified Framework for Deciphering Central Bank Communications Globally}.
\newblock
\showeprint[arxiv]{2505.17048}~[cs.CL]
\urldef\tempurl%
\url{https://arxiv.org/abs/2505.17048}
\showURL{%
\tempurl}


\bibitem[Shah et~al\mbox{.}(2022)]%
        {shah2022flue}
\bibfield{author}{\bibinfo{person}{Raj~Sanjay Shah}, \bibinfo{person}{Kunal Chawla}, \bibinfo{person}{Dheeraj Eidnani}, \bibinfo{person}{Agam Shah}, \bibinfo{person}{Wendi Du}, \bibinfo{person}{Sudheer Chava}, \bibinfo{person}{Natraj Raman}, \bibinfo{person}{Charese Smiley}, \bibinfo{person}{Jiaao Chen}, {and} \bibinfo{person}{Diyi Yang}.} \bibinfo{year}{2022}\natexlab{}.
\newblock \showarticletitle{When flue meets flang: Benchmarks and large pre-trained language model for financial domain}.
\newblock \bibinfo{journal}{\emph{arXiv preprint arXiv:2211.00083}} (\bibinfo{year}{2022}).
\newblock


\bibitem[Sprenger et~al\mbox{.}(2014)]%
        {sprengertweets2014}
\bibfield{author}{\bibinfo{person}{Timm~O. Sprenger}, \bibinfo{person}{Andranik Tumasjan}, \bibinfo{person}{Philipp~G. Sandner}, {and} \bibinfo{person}{Isabell~M. Welpe}.} \bibinfo{year}{2014}\natexlab{}.
\newblock \showarticletitle{Tweets and Trades: the Information Content of Stock Microblogs}.
\newblock \bibinfo{journal}{\emph{European Financial Management}} \bibinfo{volume}{20}, \bibinfo{number}{5} (\bibinfo{year}{2014}), \bibinfo{pages}{926--957}.
\newblock
\href{https://doi.org/10.1111/j.1468-036X.2013.12007.x}{doi:\nolinkurl{10.1111/j.1468-036X.2013.12007.x}}
\showeprint{https://onlinelibrary.wiley.com/doi/pdf/10.1111/j.1468-036X.2013.12007.x}


\bibitem[Sundar et~al\mbox{.}(2021)]%
        {10.1093/jcmc/zmab010}
\bibfield{author}{\bibinfo{person}{S~Shyam Sundar}, \bibinfo{person}{Maria~D Molina}, {and} \bibinfo{person}{Eugene Cho}.} \bibinfo{year}{2021}\natexlab{}.
\newblock \showarticletitle{Seeing Is Believing: Is Video Modality More Powerful in Spreading Fake News via Online Messaging Apps?}
\newblock \bibinfo{journal}{\emph{Journal of Computer-Mediated Communication}} \bibinfo{volume}{26}, \bibinfo{number}{6} (\bibinfo{date}{08} \bibinfo{year}{2021}), \bibinfo{pages}{301--319}.
\newblock
\showISSN{1083-6101}
\href{https://doi.org/10.1093/jcmc/zmab010}{doi:\nolinkurl{10.1093/jcmc/zmab010}}
\showeprint{https://academic.oup.com/jcmc/article-pdf/26/6/301/41139661/zmab010.pdf}


\bibitem[Symbiosis and Gandhi(2024)]%
        {Symbiosis_Gandhi_2024}
\bibfield{author}{\bibinfo{person}{Aditi~Rajput Symbiosis} {and} \bibinfo{person}{Aradhana Gandhi}.} \bibinfo{year}{2024}\natexlab{}.
\newblock \showarticletitle{Finfluencer: Exploring the untapped influence of financial influencers}.
\newblock \bibinfo{journal}{\emph{2024 14th International Conference on Advanced Computer Information Technologies (ACIT)}}  \bibinfo{volume}{2024} (\bibinfo{date}{Sep} \bibinfo{year}{2024}), \bibinfo{pages}{190–196}.
\newblock
\href{https://doi.org/10.1109/acit62333.2024.10712618}{doi:\nolinkurl{10.1109/acit62333.2024.10712618}}


\bibitem[Tatarinov et~al\mbox{.}(2025)]%
        {tatarinov2025languagemodelingfuturefinance}
\bibfield{author}{\bibinfo{person}{Nikita Tatarinov}, \bibinfo{person}{Siddhant Sukhani}, \bibinfo{person}{Agam Shah}, {and} \bibinfo{person}{Sudheer Chava}.} \bibinfo{year}{2025}\natexlab{}.
\newblock \bibinfo{title}{Language Modeling for the Future of Finance: A Quantitative Survey into Metrics, Tasks, and Data Opportunities}.
\newblock
\showeprint[arxiv]{2504.07274}~[cs.CL]
\urldef\tempurl%
\url{https://arxiv.org/abs/2504.07274}
\showURL{%
\tempurl}


\bibitem[Tkachenko et~al\mbox{.}(2022)]%
        {Label-Studio}
\bibfield{author}{\bibinfo{person}{Maxim Tkachenko}, \bibinfo{person}{Mikhail Malyuk}, \bibinfo{person}{Andrey Holmanyuk}, {and} \bibinfo{person}{Nikolai Liubimov}.} \bibinfo{year}{2020-2022}\natexlab{}.
\newblock \bibinfo{title}{{Label Studio}: Data labeling software}.
\newblock
\urldef\tempurl%
\url{https://github.com/heartexlabs/label-studio}
\showURL{%
\tempurl}
\newblock
\shownote{Open source software available from https://github.com/heartexlabs/label-studio}.


\bibitem[{U.S. Securities and Exchange Commission}(2020)]%
        {SEC_RegBI}
\bibfield{author}{\bibinfo{person}{{U.S. Securities and Exchange Commission}}.} \bibinfo{year}{2020}\natexlab{}.
\newblock \bibinfo{title}{{Regulation Best Interest: A Small Entity Compliance Guide}}.
\newblock
\urldef\tempurl%
\url{https://www.sec.gov/resources-small-businesses/small-business-compliance-guides/regulation-best-interest}
\showURL{%
\tempurl}
\newblock
\shownote{[Accessed: February 2, 2025]}.


\bibitem[Warkulat and Pelster(2024)]%
        {WARKULAT2024103721}
\bibfield{author}{\bibinfo{person}{Sonja Warkulat} {and} \bibinfo{person}{Matthias Pelster}.} \bibinfo{year}{2024}\natexlab{}.
\newblock \showarticletitle{Social media attention and retail investor behavior: Evidence from r/wallstreetbets}.
\newblock \bibinfo{journal}{\emph{International Review of Financial Analysis}}  \bibinfo{volume}{96} (\bibinfo{year}{2024}), \bibinfo{pages}{103721}.
\newblock
\showISSN{1057-5219}
\href{https://doi.org/10.1016/j.irfa.2024.103721}{doi:\nolinkurl{10.1016/j.irfa.2024.103721}}


\bibitem[Wu et~al\mbox{.}(2023)]%
        {wu2023bloomberggpt}
\bibfield{author}{\bibinfo{person}{Shijie Wu}, \bibinfo{person}{Ozan Irsoy}, \bibinfo{person}{Steven Lu}, \bibinfo{person}{Vadim Dabravolski}, \bibinfo{person}{Mark Dredze}, \bibinfo{person}{Sebastian Gehrmann}, \bibinfo{person}{Prabhanjan Kambadur}, \bibinfo{person}{David Rosenberg}, {and} \bibinfo{person}{Gideon Mann}.} \bibinfo{year}{2023}\natexlab{}.
\newblock \showarticletitle{Bloomberggpt: A large language model for finance}.
\newblock \bibinfo{journal}{\emph{arXiv preprint arXiv:2303.17564}} (\bibinfo{year}{2023}).
\newblock


\bibitem[Yang et~al\mbox{.}(2023)]%
        {yang2023fingpt}
\bibfield{author}{\bibinfo{person}{Hongyang Yang}, \bibinfo{person}{Xiao-Yang Liu}, {and} \bibinfo{person}{Christina~Dan Wang}.} \bibinfo{year}{2023}\natexlab{}.
\newblock \showarticletitle{Fingpt: Open-source financial large language models}.
\newblock \bibinfo{journal}{\emph{arXiv preprint arXiv:2306.06031}} (\bibinfo{year}{2023}).
\newblock


\bibitem[Yasunaga et~al\mbox{.}(2025)]%
        {yasunaga2025multimodalrewardbenchholisticevaluation}
\bibfield{author}{\bibinfo{person}{Michihiro Yasunaga}, \bibinfo{person}{Luke Zettlemoyer}, {and} \bibinfo{person}{Marjan Ghazvininejad}.} \bibinfo{year}{2025}\natexlab{}.
\newblock \bibinfo{title}{Multimodal RewardBench: Holistic Evaluation of Reward Models for Vision Language Models}.
\newblock
\showeprint[arxiv]{2502.14191}~[cs.CV]
\urldef\tempurl%
\url{https://arxiv.org/abs/2502.14191}
\showURL{%
\tempurl}


\bibitem[Yousefi et~al\mbox{.}(2024)]%
        {10.1145/3665026.3665039}
\bibfield{author}{\bibinfo{person}{Niloofar Yousefi}, \bibinfo{person}{Mert~Can Cakmak}, {and} \bibinfo{person}{Nitin Agarwal}.} \bibinfo{year}{2024}\natexlab{}.
\newblock \showarticletitle{Examining Multimodel Emotion Assessment and Resonance with Audience on YouTube}. In \bibinfo{booktitle}{\emph{Proceedings of the 2024 9th International Conference on Multimedia and Image Processing}} (Osaka, Japan) \emph{(\bibinfo{series}{ICMIP '24})}. \bibinfo{publisher}{Association for Computing Machinery}, \bibinfo{address}{New York, NY, USA}, \bibinfo{pages}{85–93}.
\newblock
\showISBNx{9798400716164}
\href{https://doi.org/10.1145/3665026.3665039}{doi:\nolinkurl{10.1145/3665026.3665039}}


\bibitem[Zhang et~al\mbox{.}(2023)]%
        {zhang2023positivescalingnegationimpacts}
\bibfield{author}{\bibinfo{person}{Yuhui Zhang}, \bibinfo{person}{Michihiro Yasunaga}, \bibinfo{person}{Zhengping Zhou}, \bibinfo{person}{Jeff~Z. HaoChen}, \bibinfo{person}{James Zou}, \bibinfo{person}{Percy Liang}, {and} \bibinfo{person}{Serena Yeung}.} \bibinfo{year}{2023}\natexlab{}.
\newblock \bibinfo{title}{Beyond Positive Scaling: How Negation Impacts Scaling Trends of Language Models}.
\newblock
\showeprint[arxiv]{2305.17311}~[cs.CL]
\urldef\tempurl%
\url{https://arxiv.org/abs/2305.17311}
\showURL{%
\tempurl}


\bibitem[Zhu et~al\mbox{.}(2023)]%
        {ZHU2023306}
\bibfield{author}{\bibinfo{person}{Linan Zhu}, \bibinfo{person}{Zhechao Zhu}, \bibinfo{person}{Chenwei Zhang}, \bibinfo{person}{Yifei Xu}, {and} \bibinfo{person}{Xiangjie Kong}.} \bibinfo{year}{2023}\natexlab{}.
\newblock \showarticletitle{Multimodal Sentiment Analysis Based on Fusion Methods: A Survey}.
\newblock \bibinfo{journal}{\emph{Information Fusion}}  \bibinfo{volume}{95} (\bibinfo{year}{2023}), \bibinfo{pages}{306--325}.
\newblock
\showISSN{1566-2535}
\href{https://doi.org/10.1016/j.inffus.2023.02.028}{doi:\nolinkurl{10.1016/j.inffus.2023.02.028}}


\end{thebibliography}

\newpage
\appendix

\section{What is a recommendation?}
\label{app:what-is-a-recommendation}

A \emph{recommendation} in our study refers to any explicit suggestion made by the influencer for viewers to take a specific action regarding a stock. Each recommendation encompasses the following components:

\paragraph{Action}

The \emph{action} denotes the specific advice given by the influencer. This includes advising viewers to:

\begin{itemize}
    \item \textbf{Buy}: Purchase shares of the stock.
    \item \textbf{Hold}: Retain the stock if already owned, without necessarily buying more.
    \item \textbf{Don't Buy}: Refrain from purchasing the stock.
    \item \textbf{Sell}: Sell shares of the stock currently owned.
    \item \textbf{Short Sell}: Sell shares not currently owned, intending to buy them back later at a lower price.
    \item \textbf{Unclear}: When the action is not explicitly stated.
\end{itemize}

The \emph{action source} identifies the specific point in the video where the influencer mentions the recommendation. Annotators marked the start and end timestamps of segments where stock recommendations are discussed.

\paragraph{Stock Ticker Symbol}

A stock \emph{ticker symbol} is a unique series of letters assigned to a publicly traded company's stock (e.g., ``AAPL'' for Apple Inc.). In our annotations, the ticker symbol identifies the specific company discussed by the influencer.

\paragraph{Price}

The \emph{price} refers to the specific stock price at which the influencer recommends initiating a trade. This price refers to the price at which the influencer suggests initiating the said action. For example, for a "Buy" action, the annotated price would imply the price at which the influencer is suggesting to buy the stock. Conversely, for a "Sell" action, the annotated price would imply the price at which the influencer is suggesting to sell the stock (assuming that the stock already exists in the inverstor's portfolio). 

\paragraph{Quantity}

\emph{Quantity} denotes the amount of the stock that the influencer suggests purchasing. This could be expressed as the number of shares or as a percentage allocation of an investment portfolio. For instance, an influencer might recommend investing 10\% of one's total investment funds into a particular stock. Annotating the quantity provides insight into the emphasis placed on the stock within an overall investment strategy.

\paragraph{Action Date}

\emph{Action Date} refers to the specific date on which the suggested action should take place.

\section{Conviction of Recommendation}
\begin{table}[h]
    \centering
    \caption{Evaluation of conviction based on tone, facial expressions, overall delivery, and consistency.}
    \resizebox{\columnwidth}{!}{%
    \begin{tabular}{p{0.8cm} p{2.5cm} p{3cm} p{3cm} p{2.5cm}}
        \textbf{Score} & \textbf{Tone} & \textbf{Facial Expressions} & \textbf{Overall Delivery} & \textbf{Consistency} \\
        1 & Hesitant or uncertain, frequent qualifiers. & Neutral or doubtful (furrowed brows, pursed lips). & Reserved or doubtful language. & Bold title, but the video lacks matching conviction. \\
        \midrule
        2 & Relatively confident, some qualifiers. & Moderate enthusiasm (mild smiles, slightly raised eyebrows). & Balanced and moderately positive language. & Bold title, followed by consistent confidence in the video. \\
        \midrule
        3 & Strong, assertive language without hesitation. & Enthusiastic, energetic (wide smiles, raised eyebrows). & Decisive recommendations with no qualifiers. & Title and video are strongly aligned. \\
    \end{tabular}%
    }
\end{table}

\section{Length of Video Rationale}
\label{app: length_of_video_rationale}
Given the computational limits of MLLMs, videos with shorter video duration (under 4 minutes) would have been ideal for our research. However, we discovered a scarcity of short videos on stock recommendations, and the average length of a finfluencer video in our dataset is 9 minutes. This can be attributed to two reasons. (1) Stock recommendation videos often require more time for detailed market trends, risks, and rewards analysis. (2) Short videos lack engagement in the domain compared to longer videos. To ensure we obtain sufficient number of recommendations while maintaining their relevancy, we set the maximum video length to 12 minutes. This decision aims to strike a balance between getting relevant and adequate videos while avoiding redundancy and information overload.

\section{Expert Human Annotation Interface}
\label{app:labeling_interface}

\begin{figure}[H]
    \centering
    \includegraphics[width=\columnwidth]{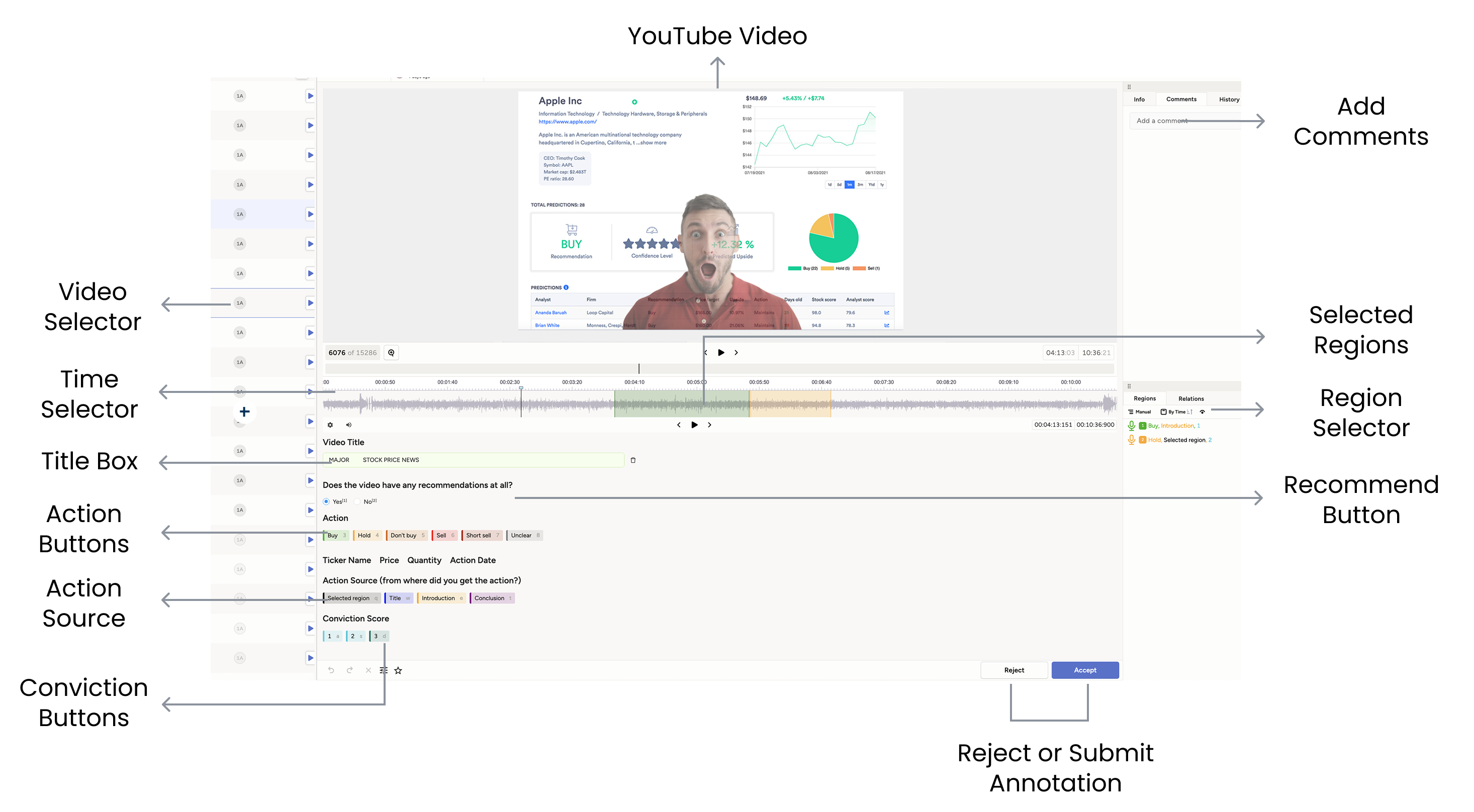}
    \caption{Annotation interface on Label Studio.}
    \label{fig:labeling_interface}
\end{figure}

\begin{table*}[H]
\caption{Conviction Score Breakdown}
\begin{tabular}{|c|p{3cm}|p{3cm}|p{3cm}|p{3cm}|}
  \hline
  \textbf{Conviction Score} & \textbf{Tone} & \textbf{Facial Expressions} & \textbf{Delivery} & \textbf{Example} \\
  \hline
  \textbf{1 (Low)} & Uncertain, hesitant, frequent pauses & Neutral or doubtful (furrowed brows, pursed lips) & Many qualifiers (e.g., “maybe,” “possibly”), closed body language (e.g., crossed arms) & “I’m not sure, but this stock might do well.” \\
  \hline
  \textbf{2 (Moderate)} & Relatively confident, steady speech & Moderate enthusiasm (mild smiles, slightly raised eyebrows) & Some qualifiers, generally positive, open body language & “I believe this stock has potential.”\\
  \hline
  \textbf{3 (High)} & Very confident, assertive, no hesitation & Enthusiastic, energetic (wide smiles, raised eyebrows)& No qualifiers, strong language, engaging body language (e.g., hand movements)&  “This stock is a must-buy.” \\
  \hline
\end{tabular}
\end{table*}

\section{\datasetname\ Metadata}
\label{app:metadata_details}
YouTube metadata encompasses all details associated with an uploaded video, excluding the video and audio content itself. The dataset includes comprehensive metadata that captures both video-level and channel-level attributes, providing a holistic view of financial content and its engagement dynamics. At the video level, attributes such as channel title, video description, and tags help contextualize the nature of the content, offering insights into the creator’s intent and the topics covered. Additional metadata like default audio language and caption availability indicate accessibility features and audience inclusivity. The video duration provides further context on content length, which can influence engagement and information density.

Engagement metrics, including likes, favorites, comments, and other audience interactions, reflect how the content resonates with viewers and the level of discussion it generates. Extracted viewer comments serve as a qualitative measure of audience sentiment and reaction. At the channel level, attributes such as channel description, channel category, total views, subscriber count, and video count help assess the creator’s influence and authority in the financial content space. These features collectively enable deeper analysis of how video characteristics, audience engagement, and creator credibility shape financial discourse on video-sharing platforms.

\section{Model Implementation Details}
\label{app: ImplementationDetails}
Model inference occured between February 1, 2025 and February 24, 2025 at a \textit{temperature} setting of 0.00 (for reproducibility). GPT models were inferenced through the OpenAI API\footnote{\url{https://openai.com/blog/openai-api}}. Gemini models were inferenced through the Gemini API\footnote{\url{https://cloud.google.com/vertex-ai/generative-ai/docs/model-reference/inference}} with all the safety settings set to block none. Anthropic models were inferenced through the Anthropic API\footnote{\url{https://github.com/anthropics/anthropic-sdk-python}}.
The open-source LLMs were inferenced through Together AI\footnote{\url{https://api.together.xyz}}. The open-source MLLMs were sampled at a rate of 0.25 frames per second and were inferenced using an A100 GPU.

\section{Financial Glossary: Definitions and Further Reading}
\label{app:financial_gloassary}

The following definitions are common financial terms used in this paper, with references to further readings.

\begin{itemize}

    \item \textbf{Annualized Return (\%):}  
    The average yearly return of an investment, allowing for a fair comparison even if different investments span different timeframes.

    \item \textbf{Backtesting}: The process of testing a trading strategy on historical data to evaluate its performance. \textit{Further reading}\footnote{\url{https://www.investopedia.com/terms/b/backtesting.asp}}

    \item \textbf{Cumulative Return (\%):}  
    The overall percentage increase (or decrease) in the investment, independent of the amount of time involved. \textit{Further reading}\footnote{\url{https://www.investopedia.com/terms/c/cumulativereturn.asp}}

    \item \textbf{Financial Portfolio}: A collection of financial investments like stocks, bonds, commodities, and other assets. \textit{Further reading}\footnote{\url{https://www.investopedia.com/terms/p/portfolio.asp}}

    \item \textbf{Index:} An index is a statistical measure that tracks the performance of a group of assets, such as stocks, bonds, or commodities, representing a specific market or sector. It serves as a benchmark for investors to gauge market trends and compare investment performance. Examples include the S\&P 500, QQQ, and NASDAQ Composite. \textit{Further reading}\footnote{\url{https://www.investopedia.com/terms/i/index.asp}}

    \item \textbf{Personal Finance Influencer ("FinFluencer")}: Individuals who share financial advice and tips, typically on social media, to help their followers make financial decisions. \textit{Further reading}\footnote{\url{https://www.investopedia.com/financial-influencers-to-know-5217608}}

    \item \textbf{PnL (\$):}  
    Profit and Loss (PnL) is how much money you gained or lost. A PnL of \$100 means a \$100 increase from your starting point.

    \item \textbf{Sharpe Ratio}: The Sharpe Ratio is used to assess the performance of different investment portfolios or financial strategies. This measures the return of an investment compared to a risk-free asset, adjusted for the investment’s volatility. A higher Sharpe ratio generally indicates better risk-adjusted performance. Mathematically, the Sharpe Ratio ($S$) is defined by the equation:

    \begin{equation}
    S = \frac{R_p - R_f}{\sigma_p}
    \label{eq:sharpe-ratio}
    \end{equation}

    where $R_p$ represents the return of the portfolio, $R_f$ represents the risk-free rate of return, and $\sigma_p$ is the standard deviation of the portfolio's excess returns. A higher $S$ indicates a better risk-adjusted return relative to the risk-free rate. A Sharpe Ratio of 0 indicates the investment's average return is exactly equal to the risk-free rate of return and a Sharpe Ratio of 1 or higher is typically considered "good", indicating that the investment's returns adequately compensate for the risk taken.

    \item \textbf{Short Sell}: The practice of selling borrowed shares in anticipation of buying them back later at a lower price, profiting from a decline in stock price. \textit{Further reading}\footnote{\url{https://www.investopedia.com/terms/s/shortselling.asp}}

    \item \textbf{Stocks}: A type of security that represents ownership in a corporation and entitles the holder to a share of the company's assets and profits. Stocks are typically traded on an exchange and have a unique ticker symbol (e.g. Apple Inc.: AAPL). This definition does not apply to other types of securities, such as exchange-traded funds (ETFs) that track a particular index or asset class, which are not considered stocks. \textit{Further reading}\footnote{\url{https://www.investopedia.com/terms/s/stock.asp}}
\end{itemize}

\section{Quantile Analysis of Individual Stock Outcomes}
\label{ap:quintile}
\begin{figure}[h]
\centering
\includegraphics[width=\columnwidth]{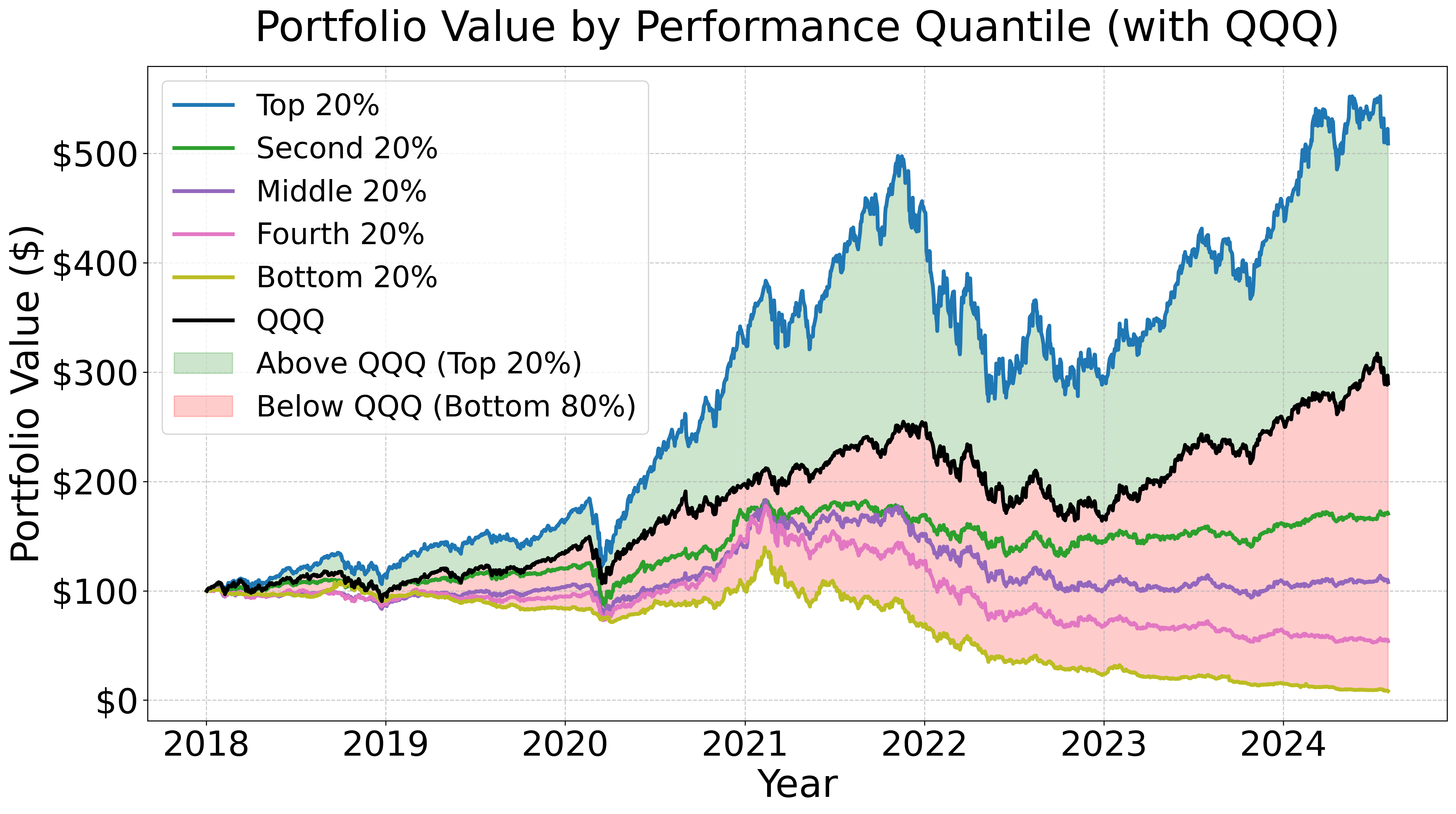}
\caption{Portfolio value by Performance Quantile (Compared with QQQ). Although a few recommendations did very well, most did not outperform a simple index fund like QQQ.}
\label{fig:quintiles}
\end{figure}

For a more detailed analysis, we simulated investing \$100 in every recommended stock and ranked them from best to worst performer, revealing that 80\% of the stocks underperformed market index strategies. To further analyze variability, we grouped all recommended stocks into quantiles (five equal groups) based on their returns. As shown in Figure \ref{fig:quintiles}, only the top 20\% stocks outperformed the QQQ, while the rest 80\% underperformed the QQQ, showing that the majority of the finfluencers stocks failed to match the performance of QQQ index. 

\section{Optimal Holding Period}

\begin{figure}[H]
\centering
\includegraphics[width=\columnwidth]{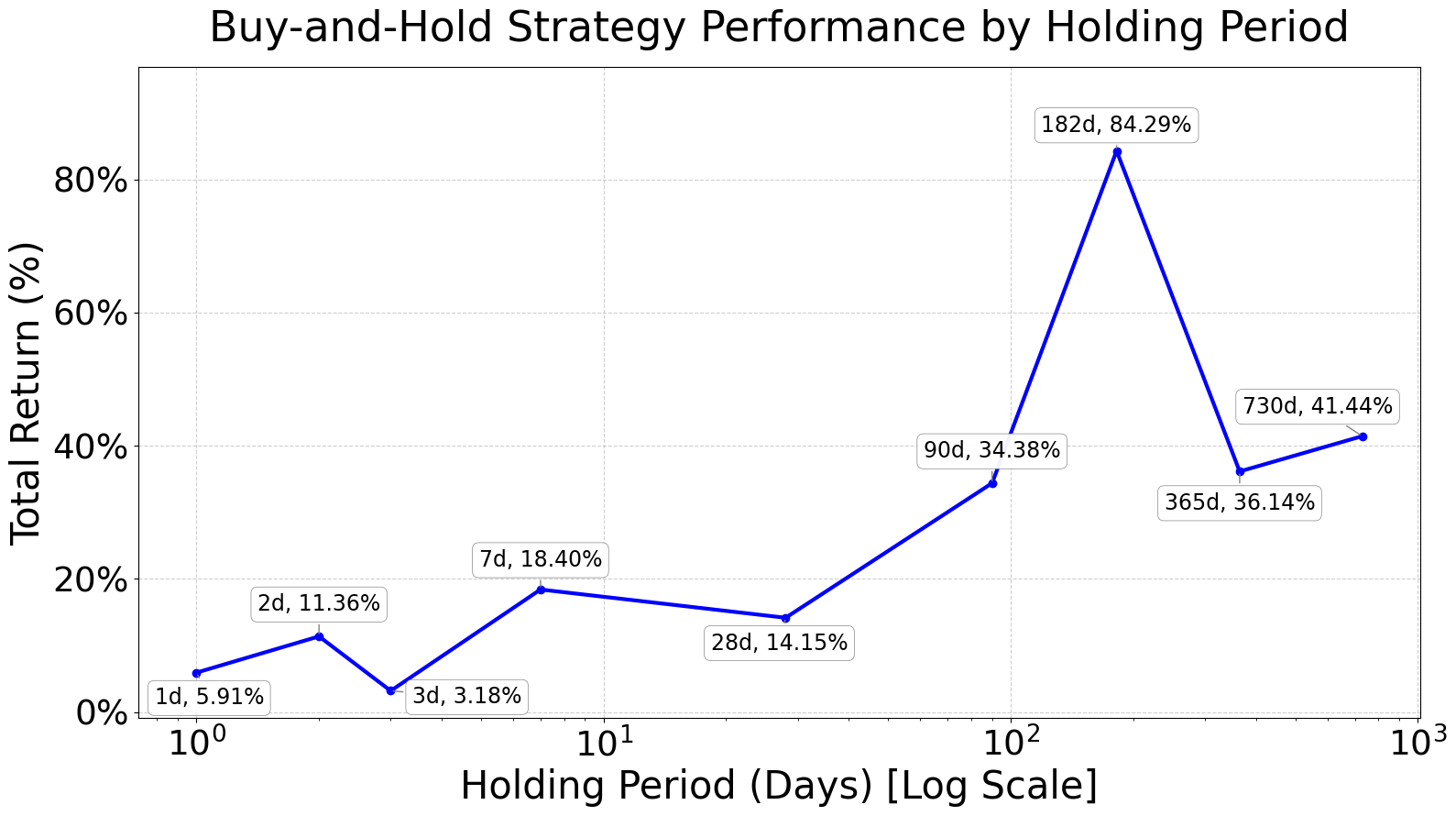}
\caption{The total return from a buy-and-hold strategy, based on the fixed holding period indicated on the x-axis. For example, at a 6-month holding period, the total return is 84.29\%. This means that between January 1, 2018, and August 1, 2024, every stock recommended for buying was held for 6 months before being sold, and repeating this strategy throughout the period resulted in a cumulative return of 84.29\%.}
\label{fig:optimal_holding_period}
\end{figure}

Figure~\ref{fig:optimal_holding_period} explores how strategy performance varies with different fixed holding periods, helping identify the most profitable time horizon for executing a Buy-and-Hold strategy.

\section{Non-Penny Stock Analysis}

To focus on more liquid and less volatile securities, we exclude penny stocks which we define as those trading below \$5 per share. This reduces the 687 total recommendations to 567. Table~\ref{tab:non_penny_strategies_performance} presents the performance of key trading strategies on this filtered subset.

\begin{table}[H]
\caption{Performance of Trading Strategies vs. Market Benchmarks, ordered by \textit{Ann.} (annual return \%). Negative returns in the Inverse YouTuber strategy occur because it involves short selling—betting that recommended stocks will fall in price.}
\label{tab:non_penny_strategies_performance}
\centering
\resizebox{\columnwidth}{!}{%
\begin{tabular}{lrrrr}
\toprule
\textbf{Strategy/Benchmark} & \textbf{Sharpe} & \textbf{PnL (\$)} & \textbf{Cumulative (\%)} & \textbf{Ann. (\%)} \\
\midrule
Inverse YouTuber    & 0.41  & 195.38  & 195.38  & 17.90  \\
QQQ              & 0.68  & 189.74  & 189.74  & 17.55  \\
Non-Penny Buy-and-Hold	& 0.58	& 122.19	& 122.19	& 12.90 \\
SPY              & 0.65  & 102.02  & 102.02  & 11.28  \\
Buy-and-Hold        & 0.46  & 84.29   & 84.29   & 9.74   \\
Non-Penny Buy-and-Hold (Weighted by Conviction) &	0.43	& 62.60	& 62.60	& 7.67 \\
Buy-and-Hold (Weighted by Conviction)   & 0.30  & 	33.35   & 	33.35  & 4.47   \\

Non-Penny Inverse YouTuber	& 0.22	& -146.40	& -146.40	& -20.97 \\

\bottomrule
\end{tabular}%
}
\end{table}

\begin{figure}[H]
\centering
\includegraphics[width=\columnwidth]{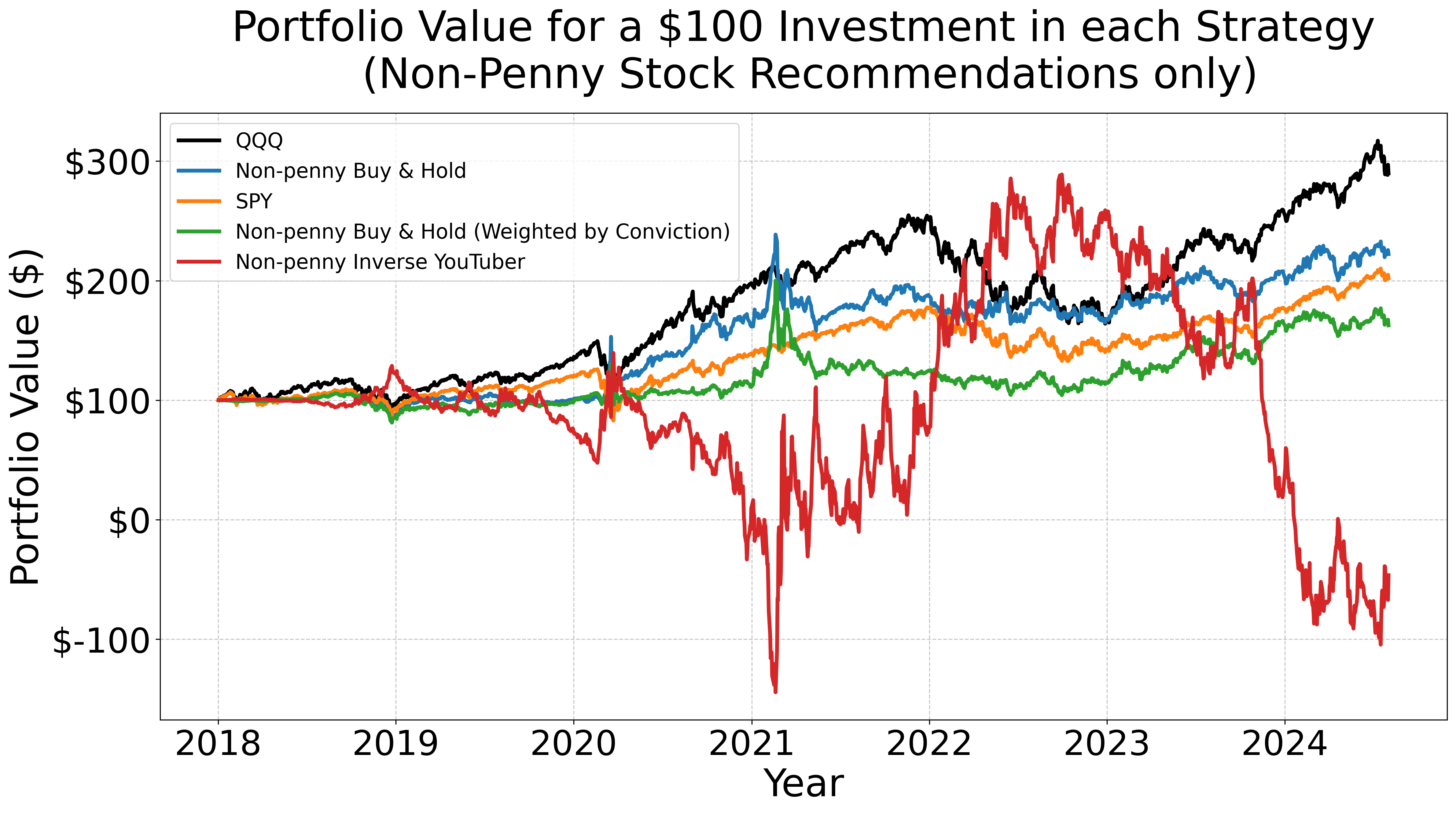}
\caption{Portfolio value based on a \$100 initial investment for non-penny stock recommendations only. QQQ outperforms all strategies, while the Non-penny buy-and-hold strategy beats the S\&P 500. The Inverse YouTuber, which was the top performer when including all recommendations, now delivers the worst returns—yielding negative returns.}
\label{fig:non_penny_strategies_performance}
\end{figure}

Figure~\ref{fig:non_penny_strategies_performance} is a non-penny stock version of Figure \ref{fig:strategies_performance}. It
provides a time series view of the trading strategies/benchmarks.

\begin{figure}[H]
\centering
\includegraphics[width=\columnwidth]{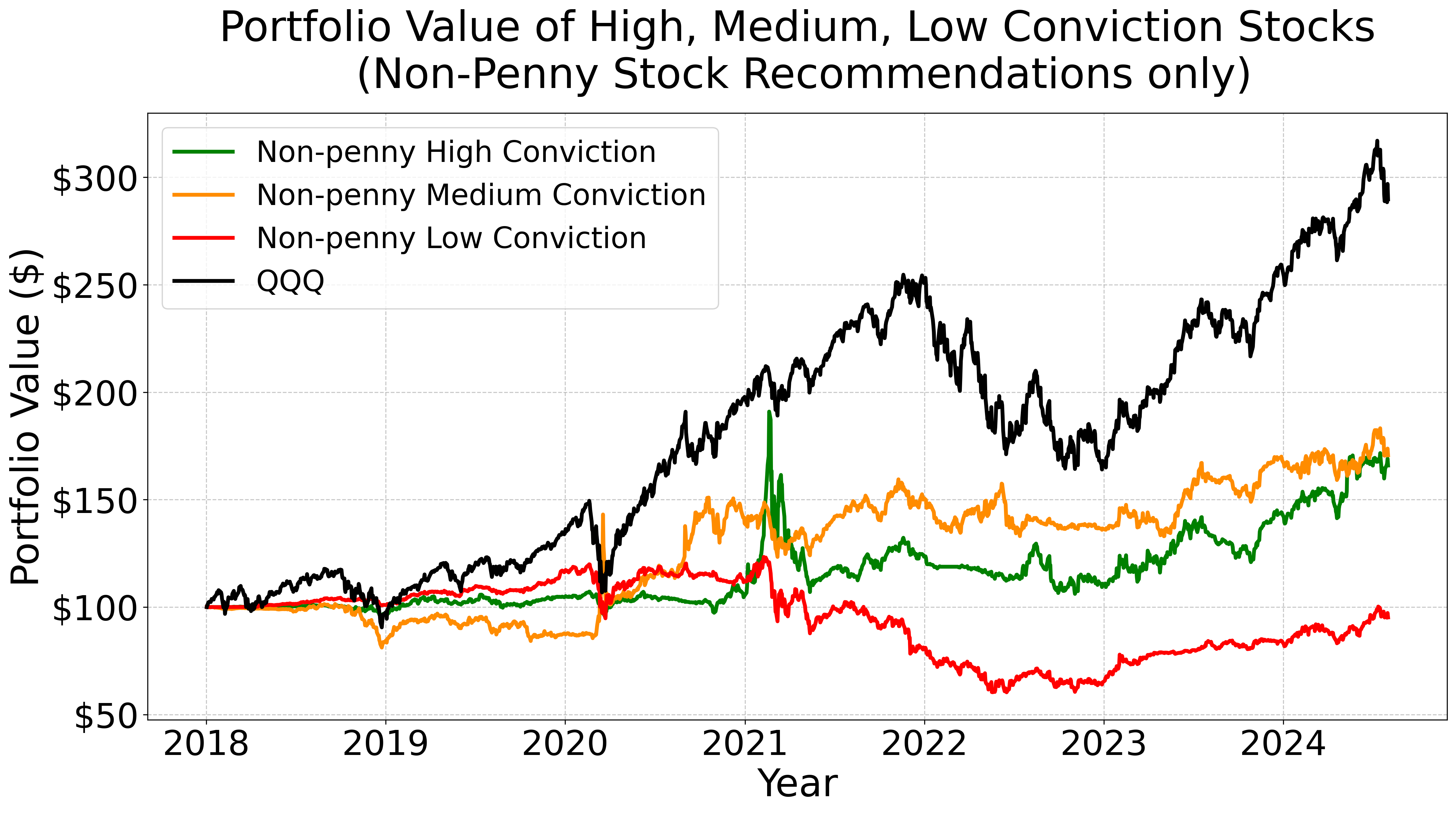}
\caption{Portfolio value based on a \$100 initial investment for non-penny stock High, Medium, and Low Conviction stocks. Medium Conviction marginally outperforms High Conviction by 4.75\%.}
\label{fig:non_penny_convictionbuckets}
\end{figure}

Figure~\ref{fig:non_penny_convictionbuckets} is a non-penny stock version of Figure \ref{fig:convictionbuckets}. It
provides a time series view of the portfolio values of high, medium, and low conviction stocks. 

Figure~\ref{fig:optimal_holding_period} explores how strategy performance varies with different fixed holding periods, helping identify the most profitable time horizon for executing a Buy and Hold strategy.

\clearpage
\onecolumn
\section{Finfluencer Background}
\label{sec:finfluencer_background}

Table~\ref{tab:channel_summary} summarizes key background details of the YouTube finfluencers in our dataset, including their self-described expertise, total views, and whether they have endorsements. URLs in self-described expertise  are redacted.

\small
\begin{longtable}{p{2.5cm} p{1.5cm} p{1cm} p{8cm} p{2cm}}
\caption{Finfluencer background based on their YouTube channel descriptions (as of April 7, 2025). Subscribers refer to the total number of people subscribed to the channel. Views are the total number of channel views, as reported by YouTube under the "More Info" section of the creator’s profile. Endorsements include paid promotions for external products/services and self-owned offerings such as courses and tools.} \label{tab:channel_summary} \\
\toprule
\textbf{Channel Name} & \textbf{Subscribers} & \textbf{Views} & \textbf{Self-Described Expertise} & \textbf{Endorsements} \\
\midrule
\endfirsthead

\toprule
\textbf{Channel Name} & \textbf{Subscribers} & \textbf{Views} & \textbf{Self Described Expertise} & \textbf{Endorsements} \\
\midrule
\endhead

\multicolumn{5}{r}{\textit{Continued on next page}} \\
\endfoot

\bottomrule
\endlastfoot

Let's Talk Money! with Joseph Hogue, CFA & 
694K &
47M & 
Welcome to your chance to create the financial future you deserve. I spent more than a decade in stock analysis for private wealth management and venture capital but I love the face-to-face interaction we get here on YouTube. I pride myself on professional analysis you won't see anywhere else on YouTube! 

Joseph Hogue is a financial expert and investment analyst. After serving in the Marine Corps, he started his career investing in real estate before becoming an investment analyst for some of the largest private investors. He's appeared on Bloomberg and on CNBC as an investment expert and has published ten books on investing. He holds a master's degree and the Chartered Financial Analyst (CFA) designation. Now he helps investors reach their financial goals and invest in the stock market with some of the same advice he used when working for the rich.
&
Yes \\
\midrule
FASTgraphs & 
147K & 
15M & 
The historical F.A.S.T. Graphs research tool provides a clear historical perspective of the company's normal operating results and prices or valuations. Their primary purpose is to illustrate the strong correlation and functional relationship between earnings and market price (in the long run). These graphs capture volumes of fundamental data at a glance. However, it's important to understand that they are only "tools to think with" and should only represent the starting point to more extensive fundamental research. On the other hand, in an instant, these powerful tools can tell the user more about the success of the businesses behind the companies they are analyzing than any other research tool available. &
Yes \\
\midrule
Ryne Williams & 
79.7K &
12M & 
This channel is all about reaching financial freedom with dividend investing. Here, I am documenting my investing journey, and hope to inspire and motivate others to start investing for themselves. I want to show that if I can do this, then so can each and every one of you. &
Yes \\
\midrule
Fin Tek & 
122K &
8M & 
Applying an Engineering Mindset to Personal Finance and Investing &
Yes \\
\midrule
Value Investing with Sven Carlin, Ph.D. &
252K &
33M & 
Stock market investing is not easy but if you apply a little bit of common sense, it can be much easier. 

Helping people to make smarter financial decisions is the mission of this channel. 

You can make better financial decisions by:

Having the right investing mindset (we do not speculate and hope - we see how the risk and reward fits our investment goals).

Doing good analyses (earnings and cash flows alongside a margin of safety is what makes a stock portfolio grow over the long-term)

Enjoy!

Disclaimer: All videos are provided for informational purposes only. Nothing contained herein should be construed as an offer, solicitation, or recommendation to buy or sell any investment or security, or to provide you with an investment strategy. Nor is this intended to be relied upon as the basis for making any purchase, sale or investment decision regarding any security. Rather, this merely expresses my opinion, which is based on information obtained from sources believed to be accurate. &
Yes \\
\midrule
Learn to Invest - Investors Grow &
286K &
20M & 
[link redacted] is an investing education website designed to simplify the world of investing; getting us all closer to our goal of achieving financial freedom. &
Yes \\
\midrule
UNRIVALED INVESTING &
105K &
9M &
WELCOME: This is a "NO HYPE, MISSION FOCUSED" channel trying to find you exceptional companies \& UNRIVALED investments!  I'm looking for stocks that have the potential to go up hundreds or even thousand's of percent overtime.  For expert investment insights, my personal financial journey \& real-money portfolio updates check out: 

[link redacted]

COMMUNITY: We also have a community of like-minded investors on our exclusive Discord server, available only for ANNUAL subscribers. 

BACKGROUND: Before launching UNRIVALED, I was lucky enough to have a variety of opportunities including CFO of a start-up that grew to \$40mn in sales \& working directly with the founder of a multi-billion dollar long-short hedge fund. More on my background here: [link redacted]

Disclaimer: [link redacted]
  &
Yes
\\
\midrule
Rational Investing with Cameron Stewart, CFA &
65.7K &
4M &
Weekly stock reviews of under valued cash flowing stocks with but high free cash flow yield and capital appreciation for the long term investor.  - Not Financial Advice 

Mr. Stewart has spend nearly 20 years in the finance industry and Wall Street, he has raised more than \$6 Billion in capital for companies seeking debt and equity financing for growth capital, dividend recapitalizations, mergers \& acquisitions or project financing.  Mr. Stewart is currently the Chief Financial Officer (CFO) for a collection of Orangetheory Gyms owned by a Private Equity Group and actively practices the principals discussed on the show. He uses his experience valuing companies to simply explain basic financial concepts and illustrate these principals during the weekly stock reviews. 

Disclaimer

Rational Investing with Cameron Stewart, CFA and [link redacted] are Not an Investment Advisors. 

Full Disclosure Here: [link redacted]
&
Yes \\ 
\midrule
ZipTrader &
701K &
80M &
Welcome to ZipTrader! ZT's Charlie Plattus places an emphasis on day-trading, swing trading, and long term investment strategies. We study price action reactions related to news as well as a focus on technical indications on up \& down trends. We strive to post the most informational and easy to understand clips on how to trade in today's volatile market. Our goal is to push our followers to develop their abilities and confidence in each and every trade.  

DISCLAIMER: All of ZipTrader, our trades, reflections, strategies, and news coverage are based on our opinions alone and are only for entertainment purposes. These are Charlie's opinions, not investment/financial/legal advice. Past performance is not a predictor of future results. This is not personalized but rather general educational and informational material. Do your own due diligence and/or consult a registered financial advisor before taking any positions. &
Yes \\
\midrule
Everything Money &
297K &
51M &
Everything Money is a disciplined investment education YouTube channel that teaches how to help build long-term wealth through stocks, real estate, and business development. Paul Gabrail is a disciplined investor who loves teaching the proper mindset, emotions, and process it takes to be a value investor. Mostafa Hussein shares his strategies that he utilizes for options trading, chart trading, and deeper dives into finding Value Stocks. 

Learn from us on how to study companies to invest in, examine real estate deals, and how to excel your business with development strategies. Learn from Paul on how he operates over 1000 units of real estate across the country. No fluff, no gimmicks, no bullsh*t....just careful, calculated investment strategies. Join our community! &
Yes \\
\midrule
Mark Roussin, CPA &
112K &
9M &
Welcome to my channel! If you are looking to boost your financial knowledge and learn about stocks, you have come to the right place!

My name is Mark -- I am a Certified Public Accountant who has been investing for nearly 15 years. I have been an active CPA in the state of California for over a decade now. I am also the founder of Roussin Financial ([link redacted]) which is my own financial business created with the intention of helping investors enhance their financial literacy. I have performed 1 on 1 coaching for a number of clients in the topics of both Personal Finances and Investing.

My goal with this YouTube channel is to spread Financial Literacy for those looking to learn. I will be discussing stock picks, investing strategies, market news, and other financial topics which are intended to be for information purposes only.

If these topics interest you, please hit that SUBSCRIBE button and let's get investing! &
Yes \\
\midrule
Finding Alpha &
21.4K &
1M & 
Welcome to the channel! Stock market investing is not easy, but here we aim to make it easier by providing high quality Analysis and Research on various publicly traded Stocks. 
Our aim, as the name suggests, is to beat the market consistently!

Subscribe to keep up to date with all the latest Stock Market News, Analysis and Valuations. &
Yes \\
\midrule
StockCharts TV &
128K &
16M &
Welcome to our channel, where we're on a relentless pursuit to help you achieve your financial goals. 

Here at StockCharts, we believe in the transformative power of knowledge, and that's why we're committed to informing, educating, and empowering traders and investors like you. Whether you're a seasoned pro or just starting out, our content is designed to equip you with the tools and insights you need to thrive in today's dynamic financial landscape.

Founded in 1999 by Chip Anderson, [link redacted] was born out of a passion for data visualization. By using technology to help investors visualize financial data, we allow our users to better analyze the markets, monitor and manage their portfolios, find promising new stocks and funds to buy, and ultimately make intelligent, well-timed investment decisions. &
Yes \\
\midrule
Felix \& Friends (Goat Academy) &
264K &
31M &
CHANNEL MISSION
The education system has one purpose.  To create the next generation of worker bees for large corporates.  

That is why we can go through school, college, even MBAs and come out without a financial education.

You deserve financial and time freedom - not 40 years of 9-5, with an insufficient pension at the end.

The only way to get there is to master managing your money.  

Stop relying on 1 salary.  

Build new income streams.  

Make your money work for you.

THE MISSION OF THIS COMMUNITY IS TO MAKE A MILLION PEOPLE FINANCIALLY FREE.

Keep motivated.

Connect you with like minded people.

WHO ARE FELIX \& WINSTON?
Felix Prehn is an economist, banker \& lawyer.  Felix and his adopted golden retriever, Winston, share their 20+ years experience of investing.

Felix lost 50\% of his first investment.  While the bank who sold it to him made 7\%+

It took a major back injury for Felix to quit the rat race.

Thus motivated, he got time and financial freedom. &
Yes \\
\midrule
BWB - Business With Brian &
211K &
11M &
After a full career at corporations such as Target and Amazon, I chose to retire at 46.  I've dedicated this channel to help educate others on how to be successful at personal finance and business. &
Yes \\
\midrule
Ale's World of Stocks &
151K &
14M &
Welcome to my world of Stocks!!! Where I share my thoughts and research on all things related to the Stock Market, Investing, Money/Currencies, and other Financial related topics. Thanks for watching and please subscribe!!! :)

My Gaming channel is called "Ale's World of Gaming" and you can find it at the following link:
[link redacted]

Please take all of my videos as entertainment, as my own opinion, and at your own risk.
I am not telling anyone how to spend or invest their money. &
Yes \\
\midrule
Mr. FIRED Up Wealth &
83.7K &
5M &
My goal is to help YOU be a better investor \& achieve financial independence!

I'm a self-made millionaire with an MBA and 25+ years of trading \& investing experience. I teach fundamentals, technicals, and everything in between as your personal finance coach. From personal finance to portfolio management \& financial freedom!

\includegraphics[height=0.75em]{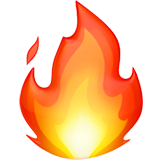} FIRED Up Wealth \includegraphics[height=0.75em]{images/Figures/fire.png} is about outperforming the stock market to achieve financial freedom \& enjoy life. This community focuses on growth, disruptive technology, barbell balance \& long-term investing. We use growth at a young age to outperform, \& we sell that outperformance to buy blue chip dividend stocks to build passive income until we have enough to be financially free \& live off dividends without selling our assets. 

\includegraphics[height=0.75em]{images/Figures/fire.png} Stock Market Investing 
\includegraphics[height=0.75em]{images/Figures/fire.png} Disruptive Technology 
\includegraphics[height=0.75em]{images/Figures/fire.png} Growth Investing 
\includegraphics[height=0.75em]{images/Figures/fire.png} Dividend Growth Investing (DGI \& DGIF)
\includegraphics[height=0.75em]{images/Figures/fire.png} Financial Independence

The info provided is for informational purposes only and should not be considered legal or financial advice. &
Yes \\
\midrule
Long Term Mindset &
167K &
6M &
\includegraphics[height=0.75em]{images/Figures/fire.png} Subscribe To Learn About Accounting \& Investing

Brian Feroldi is a financial educator, YouTuber, and author. He has written over 3,000 articles on stocks, investing, and personal finance for the Motley Fool. Brian’s best-selling book "Why Does The Stock Market Go Up?" was published in 2022. It was written to explain how the stock market works in plain English.

Book: [link redacted]

Brian Stoffel is a teacher with more than a decade of investing experience. He has written more than 4,000 articles for The Motley Fool. Brian plans his life and his investments around “antifragile” principles. &
Yes \\
\midrule
Stock Moe &
702K &
101M &
The Stock Moe YouTube channel brings the best financial education to YT. I share financial \& economic research, cryptocurrency \& stock news, XRP updates, info about DOGE stimulus checks (DOGE refund checks) \& more. I'm a former educator of high school \& college level classes. I have won 2 national championships \& 15 state championships as a Stock Market Game coach. I was a licensed stockbroker \& financial advisor with my series 7, 63, \& 65 before that. The channel which reviews these topics: Stocks, Stock Price Predictions, Cryptocurrency, Crypto, XRP, Ethereum, Inflation, Fed, Economics, Stimulus Checks, DOGE refund checks, Financial news, \& More!  Be sure to subscribe \& throw a thumbs up my way. I also go over the Stock Moe Patreon including a Stock Moe Patreon Review about the Stock Moe Discord. The Stock Moe Discord has thousands of members who are all working toward financial freedom. Get the Stock Moe Discord by signing up for the Stock Moe Patreon or become a Channel Member. &
Yes \\
\midrule
Financial Education &
846K &
127M &
My name is Jeremy Lefebvre and I created the Financial Education Channel as somewhere people from all backgrounds, countries etc can come and learn about Investing, Personal Finance and entrepreneurship!

Apply to join my private Stock group \& Wealth Group with this link

FAQ
*What do you do each day?
Now a days I focus most of my time on finding the next home run stocks, running my stock portfolios, help my private stock group members reach 6 figure, 7 figure and 8 figure milestones, and lastly record Youtube videos! I use to have a real estate marketing company before youtube took off. Before that I was a manager for a company named Quiktrip.

*Wife Kids?
Yes I have a beautiful wife and 2 awesome little boys

*When did you start the Financial Education channel?
I started in  2016

*Any advice to retail investors?
Do Research and watch my videos on investing! &
Yes \\
\midrule
Morningstar, Inc. &
104K &
15M &
We are a leading provider of independent investment research. Our mission is to empower investor success.  [link redacted] &
Yes \\
\midrule
The StockWatch &
80.1K &
4M &
Welcome in! I'm a stock market trader and investor that makes videos on trading/investment ideas, information, and opinions. I use primarily technical analysis along with some fundamental analysis and social research methods for each trading setup. For entertainment purposes only. 
Owned by: The StockWatch LLC. &
Yes

\end{longtable}

\end{document}